\documentclass[letterpaper,journal]{IEEEtran}
\usepackage{amsmath,amsfonts,amsthm}
\usepackage{algorithm, algpseudocode}
\usepackage{array}
\usepackage[caption=false,font=normalsize,labelfont=sf,textfont=sf]{subfig}
\usepackage{textcomp}
\usepackage{stfloats}
\usepackage{url}
\usepackage{verbatim}
\usepackage{graphicx}
\usepackage{multirow}
\usepackage{cite}
\usepackage{adjustbox}
\usepackage{booktabs}

\newtheorem{definition}{Definition}
\newtheorem{example}{Example}

\begin{document}

\title{Unified Neural Backdoor Removal with Only Few Clean Samples through Unlearning and Relearning}
\author{Nay Myat Min, Long H. Pham, and Jun Sun

\markboth{IEEE Transactions on Information Forensics and Security,~Vol.~XX, No.~XX, XX~2024}%
{Shell \MakeLowercase{\textit{et al.}}: A Sample Article Using IEEEtran.cls for IEEE Journals}

\thanks{The authors are with the School of Computing and Information Systems, Singapore Management University, Singapore. E-mails: myatmin.nay.2022@phdcs.smu.edu.sg, hlpham@smu.edu.sg, junsun@smu.edu.sg.}}

\maketitle

\begin{abstract}
Deep neural networks have achieved remarkable success across various applications; however, their vulnerability to backdoor attacks poses severe security risks---especially in situations where only a limited set of clean samples is available for defense. In this work, we address this critical challenge by proposing \textbf{ULRL} (\underline{U}n\underline{L}earn and \underline{R}e\underline{L}earn for backdoor removal), a novel two-phase approach for comprehensive backdoor removal. Our method first employs an \emph{unlearning} phase, in which the network's loss is intentionally maximized on a small clean dataset to expose neurons that are excessively sensitive to backdoor triggers. Subsequently, in the \emph{relearning} phase, these suspicious neurons are recalibrated using targeted reinitialization and cosine similarity regularization, effectively neutralizing backdoor influences while preserving the model's performance on benign data. Extensive experiments with 12 backdoor types on multiple datasets (CIFAR-10, CIFAR-100, GTSRB, and Tiny-ImageNet) and architectures (PreAct-ResNet18, VGG19-BN, and ViT-B-16) demonstrate that ULRL significantly reduces the attack success rate without compromising clean accuracy---even when only 1\% of clean data is used for defense.
\end{abstract}

\begin{IEEEkeywords}
Deep Neural Network, Backdoor, Mitigation.
\end{IEEEkeywords}

\section{Introduction}
\IEEEPARstart{D}{eep} Neural Networks (DNNs) have achieved remarkable success in applications such as image classification, object detection, and natural language processing. However, their deployment in mission-critical systems raises serious security concerns, particularly due to backdoor attacks, where hidden triggers embedded during training cause malicious behavior~\cite{tian2019badnet, xin2017targeted}. Backdoor attacks can be introduced via data poisoning~\cite{liu2020reflection}, neuron hijacking~\cite{yin2018trojaning}, or biased training~\cite{liu2021exray}. With the rise of AI-as-a-Service~\cite{huggingface} and increased reliance on pre-trained models and outsourced training, the risks associated with backdoor attacks have become even more pronounced. Despite considerable defense efforts, existing methods still face critical limitations. 

Defenses typically fall into detection and mitigation categories. Detection approaches include inversion-based methods~\cite{wang2019neural}, statistical anomaly detection~\cite{DBLP:journals/corr/abs-1811-00636}, and feature analysis techniques that identify abnormal neuron activations. Mitigation strategies aim to neutralize backdoors through input purification~\cite{DBLP:journals/corr/abs-1908-03369}, knowledge distillation~\cite{DBLP:journals/corr/abs-2101-05930}, and adversarial unlearning~\cite{DBLP:journals/corr/abs-2110-03735}, as well as methods based on neuron pruning or retraining~\cite{DBLP:journals/corr/abs-1805-12185, li2023reconstructive, karim2023efficient}. More recent methods, such as Feature Shift Tuning (FST)~\cite{min2023towards} and Adversarial Neuron Perturbation (ANP)~\cite{NEURIPS2021_8cbe9ce2}, attempt to recalibrate model parameters or perturb neuron activations to counteract backdoor triggers. However, these techniques often depend on large clean datasets, incur significant computational overhead, or involve trade-offs that degrade the model's performance on benign data.

Many of these defenses require complex modifications and substantial computational resources, making them impractical in resource-constrained environments. For instance, neuron pruning~\cite{li2021rethinking, DBLP:journals/corr/abs-1811-00636, DBLP:journals/corr/abs-1811-03728, Liu_2022_CVPR} demands iterative evaluations to selectively remove neurons linked to backdoors, with success that is highly dependent on the model architecture and carefully tuned pruning thresholds. Similarly, reverse-engineering methods~\cite{wang2019neural, DBLP:journals/corr/abs-2006-05646, DBLP:journals/corr/abs-1902-06531, taog2022better} are computationally expensive due to per-class optimization, and they struggle against sophisticated or semantic backdoors~\cite{guo2019tabor, bagdasaryan2019backdoor, liu2020reflection, nguyen2021wanet}. 

In contrast to these existing defenses, our proposed method, \textbf{ULRL} (\underline{U}n\underline{L}earn and \underline{R}e\underline{L}earn for backdoor removal), introduces a fundamentally novel two-phase strategy for comprehensive backdoor removal. In the \emph{unlearning} phase, we intentionally maximize the loss on a small clean dataset to expose neurons that are highly sensitive to backdoor triggers. This counterintuitive process serves as an effective diagnostic tool for identifying compromised neurons. Subsequently, in the \emph{relearning} phase, these identified neurons are recalibrated via targeted reinitialization combined with a cosine similarity regularization term, which encourages the adjusted weight vectors to diverge from their compromised states. This dual-phase approach not only effectively neutralizes backdoor influences—ensuring that suspicious neurons no longer respond to malicious triggers—but also operates efficiently with only limited clean data, thereby addressing both data scarcity and computational resource constraints. Unlike broad intervention methods such as FST~\cite{min2023towards}, which require full reinitialization of classifier neurons and risk discarding valuable learned features, ULRL precisely targets and recalibrates only those neurons most affected by backdoor signals. To the best of our knowledge, ULRL is the first method to leverage this unlearning-relearning paradigm for backdoor removal, providing precise interventions that preserve the model’s functionality on benign data while significantly reducing the attack success rate.

Our approach directly addresses the gaps left by existing methods by reducing the dependency on large clean datasets and lowering computational overhead, all while maintaining high accuracy on legitimate data. Extensive experiments on multiple datasets (CIFAR-10, CIFAR-100, GTSRB, and Tiny-ImageNet) and architectures (PreAct-ResNet18, VGG19-BN, and ViT-B-16) further validate the effectiveness and efficiency of ULRL. In summary, our contributions are:

\begin{itemize}
    \item \textbf{Refined Definition:} We propose a novel definition for neural backdoors in DNNs, laying the groundwork for an effective unified backdoor removal strategy.
    \item \textbf{Novel Dual-Phase Approach:} We introduce ULRL, a defense mechanism that uniquely integrates an \emph{unlearning} phase to expose suspicious neurons with a \emph{relearning} phase that employs cosine similarity regularization to recalibrate these neurons. This dual-phase strategy is unprecedented in its approach to backdoor removal.
    \item \textbf{Data-Efficient and Practical:} Our method effectively neutralizes a broad spectrum of backdoor attacks using only a small set of clean samples, making it highly practical for resource-constrained environments.
    \item \textbf{Comprehensive Benchmarking:} Our extensive experiments demonstrate ULRL’s superiority over five state-of-the-art defenses across 12 backdoor attack types.
\end{itemize} 

The paper is organized as follows: Section~\ref{Sec:Problem} defines the problem, Section~\ref{Sec:Approach} details our proposed method, Section~\ref{Sec:Experiment} presents experimental evaluations, Section~\ref{Sec:Ablation} conducts ablation studies, Section~\ref{Sec:Related} reviews related work, Section~\ref{Sec:Limitation} discusses limitations, and Section~\ref{Sec:Conclusion} concludes the paper.

\section{Problem Definition}
\label{Sec:Problem}

\noindent \textbf{Data and Model.} We consider a K-class classification model trained on a training dataset $\mathcal{D}$, which splits into clean $\mathcal{D}_c = \left\{\left(\boldsymbol{x}_c^{(i)}, y_c^{(i)}\right)\right\}_{i=1}^{N}$ and poisoned $\mathcal{D}_b = \left\{\left(\boldsymbol{x}_b^{(j)}, y_b^{(j)}\right)\right\}_{j=1}^{M}$ subsets. $\mathcal{D}_c$ contains $N$ clean feature vectors $\boldsymbol{x}_c^{(i)}$ with their labels $y_c^{(i)}$, while $\mathcal{D}_b$ comprises $M$ modified feature vectors $\boldsymbol{x}_b^{(j)}$ with malicious intent and corresponding labels $y_b^{(j)}$. For defense, we assume access to a small subset of $\mathcal{D}_c$ ($\mathcal{D}_d$).

\noindent \textbf{Neural Networks.} We focus on convolutional neural networks (CNNs), which map input space $\mathbb{R}^p$ to output space $\mathbb{R}^q$, denoted as $\mathcal{N}:\mathbb{R}^p \rightarrow \mathbb{R}^q$. CNNs consist of an input layer, convolutional layers for feature extraction ($\phi$), pooling layers for dimensionality reduction, and an output layer for classification ($f$). Convolutional layers apply filters to capture hierarchical patterns, followed by a non-linear activation function $\sigma$ (e.g., ReLU) to enable the learning of complex features.

\noindent \textbf{Feature Extraction and Classification.} The feature extraction function, $\phi(\theta; x): \mathbb{R}^p \rightarrow \phi(x)$, maps clean $\boldsymbol{x}_c$ and poisoned $\boldsymbol{x}_b$ inputs into a feature space $\phi(\boldsymbol{x})$. The linear classification function $\boldsymbol{f}(\boldsymbol{w}; \boldsymbol{x}) = \boldsymbol{w}^T \phi(\boldsymbol{x})$ then projects these features into the output space $\mathbb{R}^q$.

\noindent \textbf{Backdoor Attack.} A backdoor attack involves training a model with embedded backdoors and can be viewed as a dual-task learning problem~\cite{li2023reconstructive}. This requires optimizing the model’s performance on both a clean dataset $\mathcal{D}_c$ (clean task) and a backdoor dataset $\mathcal{D}_b$ (backdoor task). The challenge is to optimize parameters $\theta$, split into subsets $\theta_c$ and $\theta_b$, and weights $\boldsymbol{w}$, split into $\boldsymbol{w}_c$ and $\boldsymbol{w}_b$, for the respective tasks. The optimization goal can be formally expressed as:
\begin{equation}
    \begin{aligned}
        \underset{\theta=\theta_c \cup \theta_b, \boldsymbol{w}=\boldsymbol{w}_c \cup \boldsymbol{w}_b}{\arg \min} \Bigg[ &\underbrace{\mathbb{E}_{(\boldsymbol{x}_c, y_c) \in \mathcal{D}_c} \mathcal{L}(\boldsymbol{f}(\boldsymbol{w}_c ; \phi(\theta_c, \boldsymbol{x}_c)), y_c)}_{\text{clean task}} \\
        & + \underbrace{\mathbb{E}_{(\boldsymbol{x}_b, y_b) \in \mathcal{D}_b} \mathcal{L}(\boldsymbol{f}(\boldsymbol{w}_b ; \phi(\theta_b, \boldsymbol{x}_b)), y_b)}_{\text{backdoor task}} \Bigg],
    \end{aligned}
    \label{backdoor}
\end{equation}
where $\mathcal{L}$ is the classification loss function, such as cross-entropy. The parameters $\theta=\theta_c \cup \theta_b$ and $\boldsymbol{w}=\boldsymbol{w}_c \cup \boldsymbol{w}_b$ reflect the independence of processing clean and backdoor inputs~\cite{Cheng_Liu_Ma_Zhang_2021}, enabling backdoors to remain hidden without degrading  performance~\cite{tian2019badnet}. However, $\theta_c$ and $\theta_b$, as well as $\boldsymbol{w}_c$ and $\boldsymbol{w}_b$, may overlap (i.e., $\theta_c \cap \theta_b \neq \emptyset$ and $\boldsymbol{w}_c \cap \boldsymbol{w}_b \neq \emptyset$), indicating shared parameters for both tasks~\cite{li2023reconstructive}. This overlap complicates securing neural networks against backdoors.

\noindent \textbf{Neural Backdoor.} Traditional backdoor definitions focus on input manipulation, neglecting more complex attacks that exploit a model's internal architecture or semantic features. We introduce a refined definition based on neural mechanisms.

\begin{definition}[Backdoored Network] \label{def:NeuralBackdoor} A network $\mathcal{N}$ harbors a backdoor iff there exists a subset of neurons $S \subset \mathcal{N}$ satisfying:
\begin{itemize}
    \item \textbf{Effectiveness:} The model predicts the target label $t$ whenever neurons in $S$ are sufficiently activated;
    \item \textbf{Feasibility:} A trigger $\delta$ exists such that a backdoor function $G(\boldsymbol{x}_c, \delta)$ activates $S$, leading to predict $t$;
    \item \textbf{Shortcut-ness:} \(|S| < \epsilon_S\) where $\epsilon_S$ is a small constant, indicating the compactness of the backdoor, and $|\delta| < \epsilon_\delta$ where $\epsilon_\delta$ is a stealthiness threshold for the trigger.
\end{itemize}
\end{definition}
This definition emphasizes how backdoors exploit neural network configurations, using specific neuron subsets beyond input manipulation. It broadens the scope of attacks from input changes to nuanced semantic activations.

\noindent \textbf{Threat Model.} We consider a threat model where attackers covertly manipulate a network $\mathcal{N}$ to classify clean inputs correctly, $\mathcal{N}(\boldsymbol{x}_c) = y_c$, but misclassify poisoned inputs, $\mathcal{N}(\boldsymbol{x}_b) = y_t$, using a sophisticated trigger. The attacker controls the poisoning process without restrictions on triggers or attack complexity and aims to evade both automated and human detection. Various trigger forms can be used, enabling multifaceted attacks with diverse strategies. In this work, we assume the defender has white-box access to $\mathcal{N}$, knowing the architecture and weights but lacking control over the initial training. The defense uses a small clean dataset $\mathcal{D}_d$, free of poisoned instances, to assess and correct the model's behavior.

\noindent \textbf{Defense Objective.} We define four main defense objectives.
\begin{enumerate}
    \item \textbf{Backdoor Mitigation:} To neutralize backdoors in the neural network $\mathcal{N}$, ensuring it mitigates diverse backdoor attacks, including those with semantic triggers.
    
    \item \textbf{Preserve Performance:} To balance security with performance, ensuring that defense mitigates backdoor without sacrificing the model's accuracy on clean samples.
    
    \item \textbf{Minimize Data Dependency:} To effectively mitigate backdoors with minimal clean data, ensuring practicality in data-limited or cost-sensitive scenarios.
    
    \item \textbf{Practicality and Scalability :} To ensure easy integration and minimal computational demands, countering existing defenses' complexity and resource intensiveness. 
\end{enumerate}

\section{Our Approach}
\label{Sec:Approach}

In this section, we present our ULRL approach in detail.

\begin{figure*}
  \centering
  \includegraphics[width=0.95\linewidth]{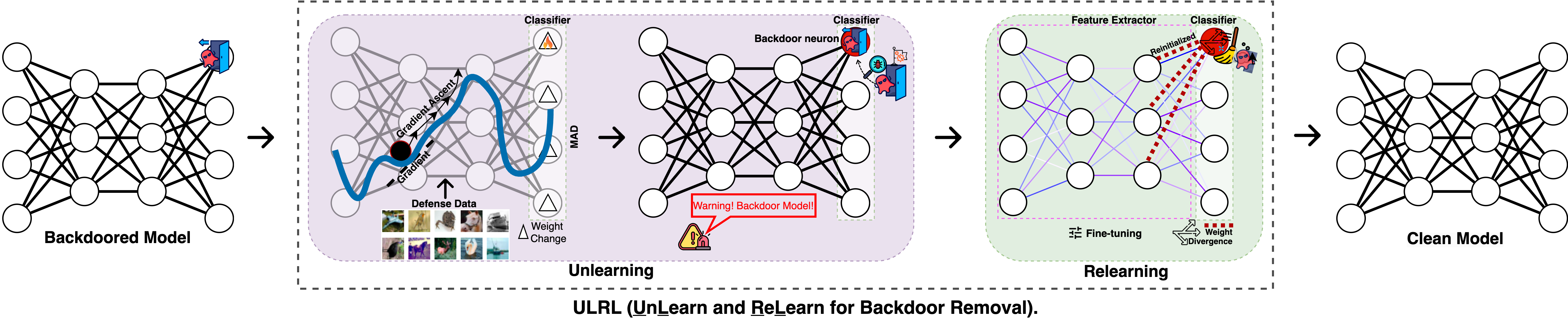}
  \caption{Overview of our proposed ULRL method. (1) A DNN model with embedded backdoors. (2) Unlearning by maximizing loss to identify suspicious neurons, applying MAD thresholds for aggregated weight changes, and Relearning by reinitialization and fine-tuning neuron weights with regularization, ensuring divergence from compromised states. (3) Purified model.}
  \label{fig:ULRL_overview}
\end{figure*}

\subsection{Overview}
Our ULRL approach addresses sophisticated backdoor attacks through two key phases, illustrated in Figure~\ref{fig:ULRL_overview} and detailed in Algorithm \ref{alg:ULRL_overview}. These phases are as follows:

\begin{enumerate}
    \item \textbf{Unlearning:} This phase challenges the conventional model training paradigms by intentionally deteriorating the performance with the defense dataset, $\mathcal{D}_d$. Instead of minimizing loss to improve accuracy, we strategically maximize loss, leveraging the insight that neurons influenced by the backdoor respond differently to this process. This phase identifies suspicious neurons by identifying abnormal weight changes during unlearning.

    \item \textbf{Relearning:} Once suspicious neurons are identified, this phase recalibrates their weights using targeted reinitialization neutralizing backdoors while preserving the model's clean performance. It is designed to enhance the model's resistance to backdoor attacks, promoting divergences in neuron weight orientations to dismantle embedded backdoor mechanisms.
\end{enumerate}

\begin{example}
\label{ex:overview}
\textit{As a running example, we use a PreAct-ResNet18 model trained on CIFAR-10, compromised by a blended backdoor attack~\cite{xin2017targeted} that subtly incorporates a near-invisible trigger into 10\% of the training data. This attack causes misclassification of inputs with the trigger while maintaining normal clean accuracy. The target class is 0. The architecture includes an initial convolution layer, four PreActBlock layers, adaptive average pooling, and a final linear layer with 10 neurons. The model achieves 93.47\% clean accuracy (C-ACC) and 99.92\% attack success rate (ASR).}
\label{eg:overview}
\end{example}

Example~\ref{ex:overview} will serve as the {\em running example} for the
rest of Section~\ref{Sec:Approach}.  
We will revisit the {\em same} compromised model in
Example~\ref{ex:unlearning} (to illustrate the unlearning phase) and again in
Example~\ref{ex:relearning} (to illustrate the relearning phase).

\subsection{Unlearning}

Our method is designed based on the hypothesis that neurons associated with backdoors will exhibit significant weight changes when subjected to unlearning with clean samples. This hypothesis is based on the understanding that due to their unique adaptation to recognize backdoor triggers, backdoor-related neurons respond differently from normal neurons when subjected to conditions that reverse their learning. Specifically, we propose that these neurons possess a unique composition of weights related to clean, backdoor, and irrelevant features. During an unlearning process, we anticipate that the weights associated with the backdoor feature will demonstrate pronounced malleability and undergo significant alterations, fundamentally due to the small number of samples used for backdoor attack, relative to that of the normal training set. This behavior is expected to starkly contrast with the stability of clean feature-related weights, aligned with the model's primary classification tasks, and the minimal changes in irrelevant weights. In the following, we first present how unlearning works and then discuss how unlearning allows us to identify these neurons associated with backdoor-related features.

\begin{algorithm}[t]
\small
\caption{ULRL (\underline{U}n\underline{L}earn and \underline{R}e\underline{L}earn)}
\begin{algorithmic}[1]
    \Statex \hspace{-\algorithmicindent} \textbf{Input:} Backdoor-ed model $f(\cdot)$
    \Statex \hspace{-\algorithmicindent} \textbf{Output:} Purified model $f'(\cdot)$
    \State \textbf{Phase 1: Unlearning}
    \State Identify suspicious neurons \Comment{Algorithm~\ref{algo:Unlearning}}
    \Statex
    \State \textbf{Phase 2: Relearning}
    \State Neutralize suspicious neurons' influence
    \State \Return Purified model $f'(\cdot)$
    \Comment{Algorithm~\ref{algo:Relearning}}
    \end{algorithmic}
    \label{alg:ULRL_overview}
\end{algorithm}

Intuitively, we use unlearning as a diagnostic tool. Unlike traditional training methods that improve model accuracy, unlearning deliberately exacerbates the model's loss for a specified task, effectively diminishing its performance. In our setting, unlearning leverages a small set of clean data $\mathcal{D}_d$ and is achieved through the following objective:
\begin{equation}
\arg \max_{\theta, \boldsymbol{w}} \mathbb{E}_{(\boldsymbol{x}_d, y_d) \in \mathcal{D}_d} \mathcal{L}(\boldsymbol{f}(\boldsymbol{w} ; \phi(\theta, \boldsymbol{x}_d)), y_d),
\label{loss_function}
\end{equation}
where $\mathcal{L}$ denotes the cross-entropy loss function, and the pairs $(\boldsymbol{x}_d, y_d)$ correspond to the clean samples and their associated labels within the dataset, $\mathcal{D}_d$. The parameters of the model after the unlearning process are indicated by $\hat{\theta}$ and $\hat{\boldsymbol{w}}$, respectively. Unlearning aims to systematically degrade the model's performance by maximizing loss on a defense dataset, targeting individual parameters for a detailed analysis of backdoor-vulnerable neurons. The unlearning phase is designed to converge by employing a clear stopping criterion: the process terminates once the model's accuracy on the clean defense dataset $\mathcal{D}_d$ falls below a predefined threshold, $CA_{\min}$. This condition prevents indefinite gradient ascent and ensures that the phase concludes when sufficient degradation has been achieved to reliably expose backdoor-related neurons.

\begin{figure}[t]
  \centering
  \includegraphics[width=0.98\linewidth]{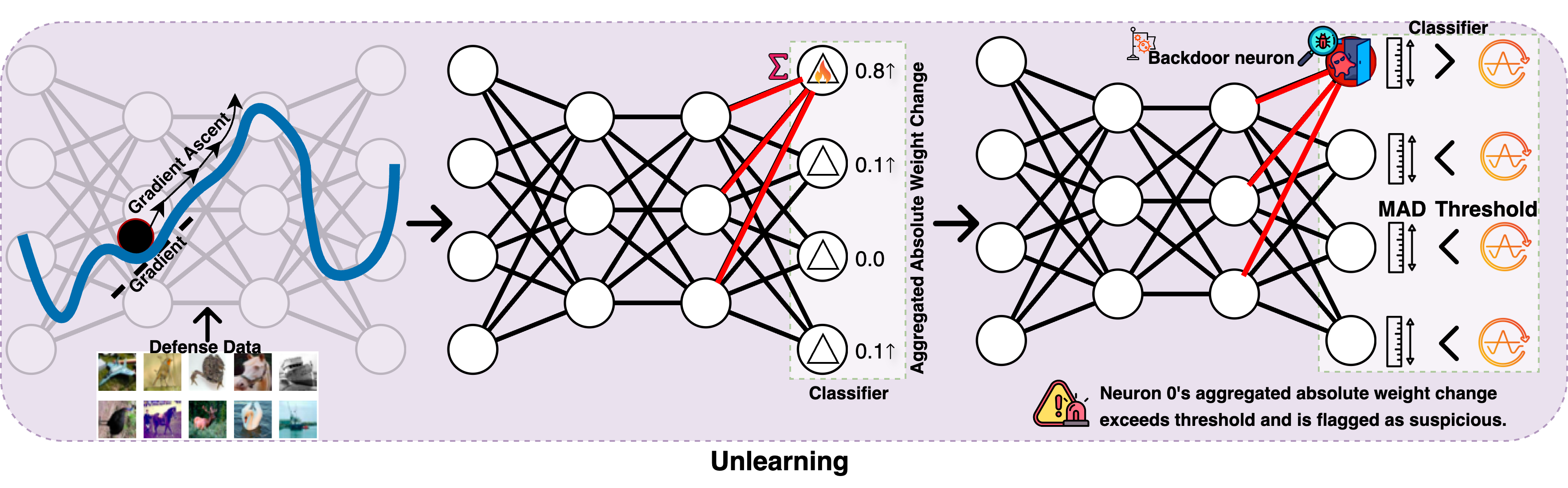}
  \caption{Details of Unlearning}
  \label{fig: Unlearning}
\end{figure}

\begin{algorithm}[t]
\small
\caption{Unlearning}
\label{algo:Unlearning}
\begin{algorithmic}[1]
\Statex \hspace{-\algorithmicindent} \textbf{Input:} Backdoor-ed model $\boldsymbol{f}(\boldsymbol{w}; \boldsymbol{\phi}(\theta; \cdot))$, Defense dataset $\mathcal{D}_d = \{ (\boldsymbol{x}_d^{(i)}, y_d^{(i)})\}_{i=1}^{N_d}$, Learning rate $\kappa$, Clean Accuracy Threshold $CA_{\min}$, MAD Threshold $\tau$, Hard Threshold HT
\Statex \hspace{-\algorithmicindent} \textbf{Output:} Set of suspicious neurons $\mathcal{S}$
\State Capture pre-unlearning classifier weights $\boldsymbol{w}_{\text{pre}} \leftarrow \boldsymbol{w}$
\Statex \hspace{-\algorithmicindent} \textbf{\#Unlearning phase}
\State Initialize $\hat{\theta} \leftarrow \theta, \hat{\boldsymbol{w}} \leftarrow \boldsymbol{w}$
\While{$CA_{f(\hat{\theta},\hat{\boldsymbol{w}})}(\mathcal{D}_d) > CA_{\min}$}
    \State Sample a mini-batch $(\boldsymbol{X}_d, \boldsymbol{Y}_d)$ from $\mathcal{D}_d$
    \State $\nabla_{\theta}, \nabla_{\boldsymbol{w}} \leftarrow \nabla_{\theta, \boldsymbol{w}} \mathcal{L}(f(\boldsymbol{X}_d; \hat{\boldsymbol{w}}, \hat{\theta}), \boldsymbol{Y}_d)$ 
    \State $\hat{\theta} \leftarrow \hat{\theta} + \kappa \nabla_{\theta}, \quad \hat{\boldsymbol{w}} \leftarrow \hat{\boldsymbol{w}} + \kappa \nabla_{\boldsymbol{w}}$ \Comment{Gradient ascent}
\EndWhile
\State Capture post-unlearning classifier weights $\boldsymbol{w}_{\text{post}} \leftarrow \hat{\boldsymbol{w}}$
\Statex \hspace{-\algorithmicindent} \textbf{\# Identification phase}
\State Compute absolute weight changes $\Delta \boldsymbol{w} = |\boldsymbol{w}_{\text{post}} - \boldsymbol{w}_{\text{pre}}|$
\State Compute absolute deviations $\Delta \boldsymbol{w}' = |\Delta \boldsymbol{w} - \text{median}(\Delta \boldsymbol{w})|$
\State Compute MAD $= \text{median}(\Delta \boldsymbol{w}')$
\State Sort in descending order $\Delta  \boldsymbol{w}^{'sorted} = \text{sort(}\Delta 
{w'}\text{)}$
\For{each $\Delta \boldsymbol{w}^{'sorted}_i$}
    \If{$\Delta \boldsymbol{w}^{'sorted}_i > \tau \times \text{MAD} \text{ \&\& } |\mathcal{S}| < \text{HT}$}
        \State Add neuron $i$ to $\mathcal{S}$ 
    \EndIf 
\EndFor

\State \textbf{return} $\mathcal{S}$ \Comment{Reload initial parameters and provide $\mathcal{S}$}
\end{algorithmic}
\end{algorithm}

Next, in Algorithm~\ref{algo:Unlearning}, we detail a method to identify neurons related to the backdoor via unlearning. First, we record the initial classifier weights, $\boldsymbol{w}_{\text{pre}}$, as a baseline (line 1). This baseline is crucial for subsequent comparisons to identify significant weight changes $\Delta \boldsymbol{w}$, revealing neurons' reactions to unlearning. During unlearning, we maximize the loss on defense data via gradient ascent until the model's performance approaches that of a random guess, meeting a minimal accuracy threshold (line 3). After unlearning, we evaluate the new weight configuration, $\boldsymbol{w}{\text{post}}$ (line 8), to isolate neurons highly sensitive to unlearning by analyzing absolute weight changes (line 9). Using MAD, a robust statistical tool~\cite{Hampel1974TheIC}, we identify neurons with substantial deviations as suspicious if their MAD scores exceed the threshold $\tau$ and the number of identified neurons is below a predefined hard threshold (HT) (lines 10-17). To prioritize the most affected neurons, we sort the neuron indices in descending order of their deviations (line 14). Finally, we revert to the original weights, $\boldsymbol{w}_{\text{pre}}$, to maintain the model's original performance (line 18).

\begin{figure}[t]
  \centering
  \includegraphics[width=0.99\linewidth]{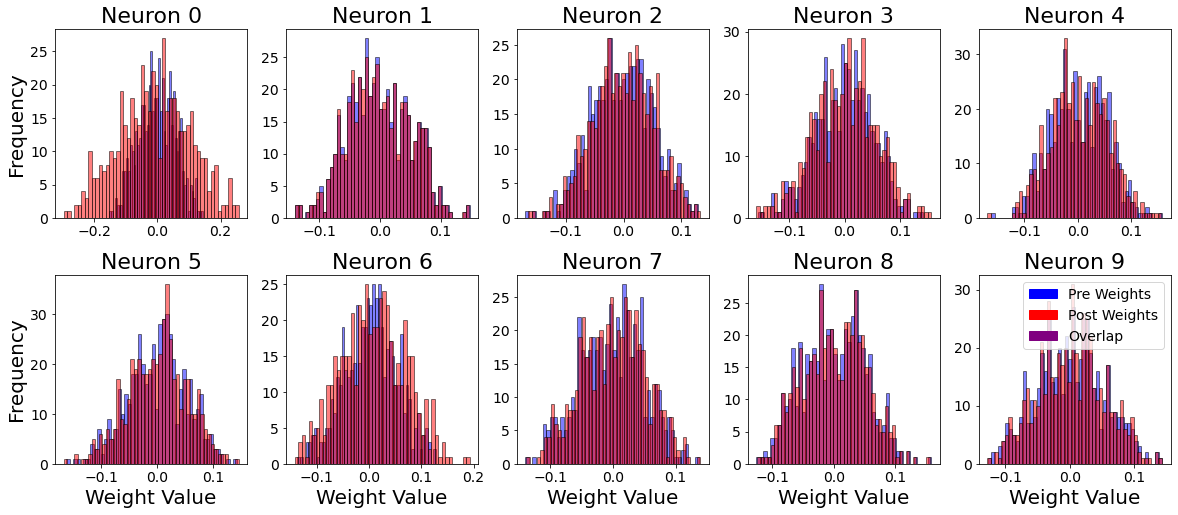}
  \caption{Neuron Weight Distribution Comparison}
  \label{fig: Dist}
\end{figure}

\begin{example}
\label{ex:unlearning}

\textit{Continuing from Example~\ref{ex:overview}, Figure~\ref{fig: Unlearning} illustrates the unlearning process. Initially, the model had 98.8\% training accuracy and 88.76\% test accuracy. Unlearning led to a pronounced decrease in training accuracy to 89.8\% and test accuracy to 27.98\% after the first epoch. The subsequent epoch dropped training accuracy to 46.6\% and test accuracy to 4.18\%, halting when training accuracy dipped below 20\%. This degradation highlighted neurons most impacted by the backdoor, with Neuron 0's standard deviation and Neuron 6's mean weight showing significant changes, suggesting their involvement in the backdoor. The weight shifts in the classifier layer, shown in Figure~\ref{fig: Dist}, flagged Neurons 0 and 6 as suspicious, with weight changes of 23.20 and 12.43, exceeding the MAD threshold of 3.5. While Neuron 6's histogram may look visually less pronounced than Neuron 0, its weight change (12.43) is still large enough to surpass MAD threshold, thereby indicating its abnormality.} No more than two neurons were reinitialized, per the hard threshold of 2.
\end{example}

\noindent \textbf{Discussion.} In the following, we analyze why backdoor neurons likely show high sensitivity to unlearning. 

Unlearning impacts DNNs, particularly at the crucial linear classifier layer, where high-level features are mapped to class labels. This process induces weight adjustments \(\Delta \boldsymbol{w}\) in DNN, revealing two trends: neurons for clean data \(\boldsymbol{f}(\boldsymbol{w}_c ; \phi(\boldsymbol{x}_c))\) experience changes $\Delta \boldsymbol{w}_c$ that impair their classification of clean inputs, while backdoor neurons \(\boldsymbol{f}(\boldsymbol{w}_b ; \phi(\boldsymbol{x}_b))\) show resilience, even improving decision making. This leads to a dynamic shift, reducing efficacy on clean data but increasing sensitivity to backdoor triggers, a counterintuitive outcome as backdoor neurons become more active and responsive post-unlearning. We analyze how unlearning affects neurons differently based on their feature associations—clean, backdoor, and irrelevant features—following unlearning. First, neurons correlated with irrelevant features, which neither aid the primary task nor trigger backdoors, show negligible changes, underscoring their peripheral role. The primary goal of the unlearning is deliberately tweaking the model's parameters to amplify the loss it incurs when evaluating clean samples. This manipulation entails the modification of the weights corresponding to clean and backdoor features. Specifically, the weights tied to clean features undergo steady reductions, diminishing their influence on the network's ability to classify correctly. Consequently, this leads to a pronounced decline in the network's capacity to discern and learn from legitimate patterns. Although the main focus is compromising performance, alterations to the weights related to backdoor features emerge as a collateral consequence of the push to maximize classification loss. 

Let us consider two cases. First, if the backdoor target is different from $y_d$, the unlearning objective can be achieved by amplifying backdoor neurons' (negative) contribution (i.e., to make the model more wrong). Second, suppose that the backdoor target is the same as $y_d$, the unlearning objective can be achieved by amplifying the contribution of backdoor neurons in a negative way (i.e., the absence of the backdoor increases the loss). In both cases, there is an incentive to modify the backdoor neurons to achieve the unlearning objective.
 
\subsection{Relearning}

After pinpointing the backdoor-related neurons, the next step is to reinitialize their weights to dismantle the embedded backdoor's influence. We use He initialization, renowned for maintaining the variance of neuron activations across layers~\cite{he2015delving}, ensuring the model's clean performance remains unaffected. For each backdoor neuron $i$ identified in the last step $\mathcal{S}$, weights are recalibrated using a zero-centered uniform distribution based on the standard deviation (lines 1-5). This approach maintains the model's learned behaviors on clean data and requires minimal retraining, offering a resource-efficient alternative to a full reset~\cite{min2023towards}.

After reinitialization, we advance our defense strategy against backdoors with relearning. Relearning, described in Algorithm~\ref{algo:Relearning}, applies a targeted regularization approach to ensure the backdoor neurons substantially diverge from potentially compromised configurations that may retain backdoor influences. This process is inspired by foundational works~\cite{kumar2022finetuning, DBLP:journals/corr/abs-1805-12185, min2023towards}, adopting an end-to-end parameter update strategy, $\{\boldsymbol{\theta},\boldsymbol{w}\}$, with a particular focus on the backdoor neurons of the classifier layer, $\mathcal{S}$. The objective of relearning is formalized as an optimization problem to minimize the cosine similarity between the updated and original weights of suspicious neurons, $\boldsymbol{w}_i$ and $\boldsymbol{w}_i^{\text{ori}}$, promoting substantial divergence:
\begin{equation}
    \begin{aligned}
        \underset{\boldsymbol{\theta}, \boldsymbol{w}}{\arg \min} \Bigg[ &\underbrace{\mathbb{E}_{(\boldsymbol{x}, \boldsymbol{y}) \sim \mathcal{D}_T}[\mathcal{L}(\boldsymbol{f}(\boldsymbol{w} ; \boldsymbol{\phi}(\boldsymbol{\theta} ; \boldsymbol{x})), \boldsymbol{y})]}_{\text{clean task performance}} \\
        & + \underbrace{\alpha \cdot \sum_{i \in \mathcal{S}} \frac{\boldsymbol{w}_i \cdot \boldsymbol{w}_i^{\text{ori}}}{\|\boldsymbol{w}_i\|_2 \|\boldsymbol{w}_i^{\text{ori}}\|_2}}_{\text{targeted weight divergence}} \Bigg],
    \end{aligned}
    \label{weight_divergence}
\end{equation}

The relearning phase employs stochastic gradient descent (SGD) with momentum to optimize a composite loss function that combines the cross-entropy loss (for clean data performance) with a cosine similarity regularization term (to ensure divergence from compromised weights). Empirically, we observe that the relearning phase converges reliably within a fixed number of epochs, as evidenced by the stabilization of both the loss and performance metrics. Although we experimented with alternative optimizers like Adam, SGD consistently delivered more stable convergence. This formulation ensures that the clean performance is maintained via the cross-entropy loss, $\mathcal{L}$, while the regularization term, scaled by the parameter $\alpha$, specifically attenuates the backdoor capabilities of the identified neurons by minimizing their cosine similarity. The $\alpha$ balances the components of the loss function, managing the trade-off between preserving accuracy and enhancing robustness against backdoors. A higher value of $\alpha$ promotes more significant shifts in feature representations. Relearning neutralizes backdoor influences through subtle but impactful adjustments in feature representations, thereby achieving a balance between security and operational efficiency.

\begin{example}
\label{ex:relearning}
\begin{figure}[t]
  \centering
  \includegraphics[width=0.98\linewidth]{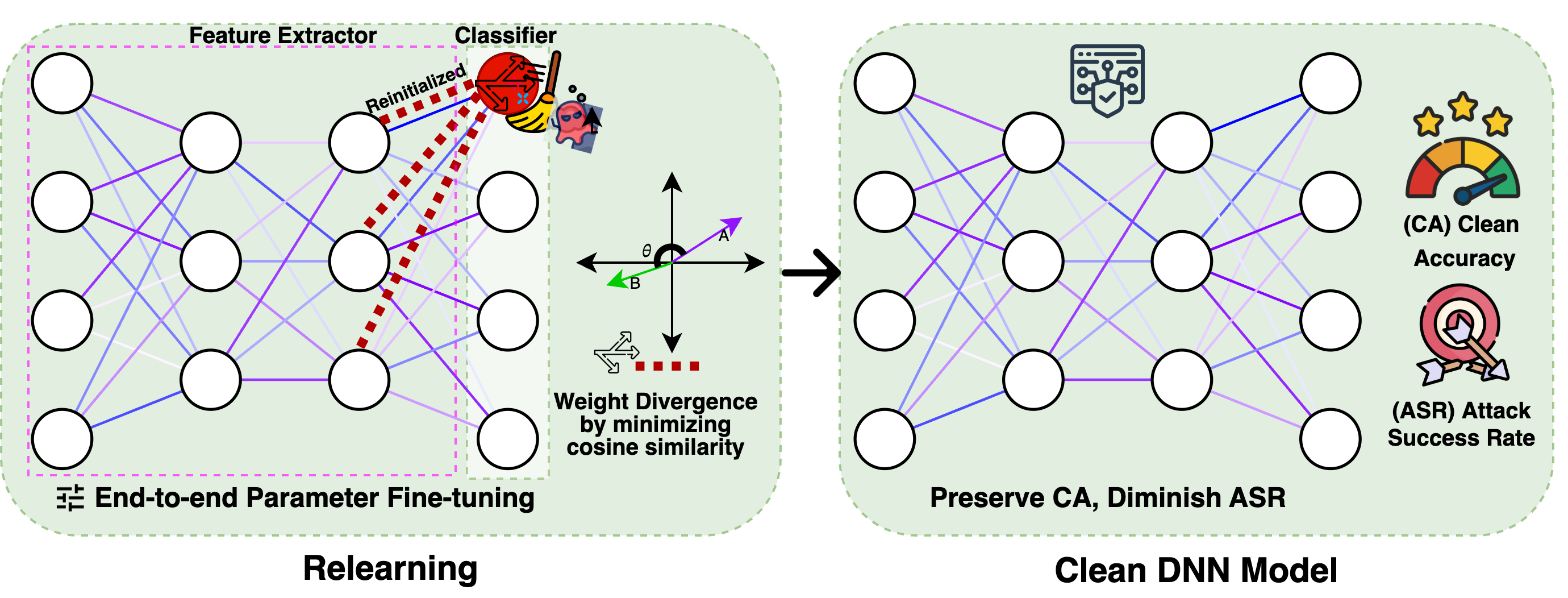}
  \caption{Details of Relearning}
  \label{fig: Relearning}
  
\end{figure}

\textit{Continuing the running example, Example~\ref{eg:overview}, Figure~\ref{fig: Relearning} illustrates the relearning process. We reinitialize the weights of suspicious neurons (0 and 6) using He initialization, impacting 2 * (512 + 1) parameters. This is highlighted for neuron 0 with red dashed lines indicating the initialized parameters. To ensure these weights diverge from their compromised states, we minimize cosine similarity, driving their orientation beyond orthogonal. This regularization is applied only to suspicious neurons, while the rest undergo standard fine-tuning. Over 20 epochs, ASR drops significantly, reaching 28.09\% after three epochs and 0.49\% by the eighth, with test accuracy stabilizing at 90.43\%. After this point, no substantial changes in ASR or test accuracy occur, suggesting that the model has reached its mitigation capacity. This demonstrates that targeting just two neurons effectively neutralizes backdoors while preserving model performance.}
\end{example}

\begin{algorithm}[t]
\small
\caption{Relearning}
\label{algo:Relearning}
\begin{algorithmic}[1]
\Statex \hspace{-\algorithmicindent} \textbf{Input:} Backdoor-ed model $\boldsymbol{f}(\boldsymbol{w}^{\text{ori}} ; \boldsymbol{\phi}(\theta; \cdot))$, Suspicious neurons $\mathcal{S}$, Defense dataset $\mathcal{D}_d = \{ (\boldsymbol{x}_d^{(i)}, y_d^{(i)})\}_{i=1}^{N_d}$, Learning rate $\eta$, Regularization factor $\alpha$
\Statex \hspace{-\algorithmicindent} \textbf{Output:} Purified model 

\Statex \hspace{-\algorithmicindent} \textbf{\# Reinitialization of Suspicious Neurons}
\State Compute He initialization STD: $\text{std} = \sqrt{\frac{2}{\text{fan}_{\text{in}}}}$
\For{each neuron $i \in \mathcal{S}$}
    \State Reinitialize weights: $\boldsymbol{w}_{i} \sim \mathcal{U}(-\text{std}, \text{std})$
    \State Reset biases: $\boldsymbol{b}_{i} = 0$
\EndFor

\Statex \hspace{-\algorithmicindent} \textbf{\# Relearning}
\For{$t = 0$ to $T-1$}
    \State Sample mini-batch $\mathcal{B}_t$ from defense dataset $\mathcal{D}_d$
    \State Calculate gradients for $\theta^t$ and $\boldsymbol{w}^t$:
    \Statex \quad\quad $\nabla_{\theta^t} \mathcal{L} = \frac{1}{|\mathcal{B}_t|} \sum_{(\boldsymbol{x}, y) \in \mathcal{B}_t} \mathcal{L}(\boldsymbol{f}(\boldsymbol{w}^t ; \boldsymbol{\phi}(\theta^t ; \boldsymbol{x})), y)$
    \Statex \quad\quad $\nabla_{\boldsymbol{w}^t} \mathcal{L} = \frac{1}{|\mathcal{B}_t|} \sum_{(\boldsymbol{x}, y) \in \mathcal{B}_t} \mathcal{L}(\boldsymbol{f}(\boldsymbol{w}^t ; \boldsymbol{\phi}(\theta^t ; \boldsymbol{x})), y)$
    \Statex \quad\quad\quad\quad\quad $ + \medspace \alpha \cdot \sum_{i \in \mathcal{S}} \frac{\boldsymbol{w}_i^t \cdot \boldsymbol{w}_i^{\text{ori}}}{\|\boldsymbol{w}_i^t\|_2 \|\boldsymbol{w}_i^{\text{ori}}\|_2}$
    \State Update model parameters
    \Statex \quad\quad $\theta^{t+1} = \theta^t - \eta \nabla_{\theta^t} \mathcal{L}$, \quad $\boldsymbol{w}^{t+1} = \boldsymbol{w}^t - \eta \nabla_{\boldsymbol{w}^t} \mathcal{L}$
\EndFor

\State \textbf{return} Purified model $\boldsymbol{f}(\boldsymbol{w}^{T} ; \boldsymbol{\phi}(\theta^{T} ; \cdot))$
\end{algorithmic}
\end{algorithm}

\noindent \textbf{Discussion.} We discuss our objective function in detail, especially choosing cosine similarity as a regularization.

The relearning objective function can be decomposed into two main components: clean task performance and targeted weight divergence. In contrast to traditional fine-tuning, which prioritizes clean task performance, our approach includes a complexity layer by adding divergence, which not only optimizes the linear classifier weight $\boldsymbol{w}$ for clean performance but also ensures that it diverges from the potentially compromised initial state ($\boldsymbol{w}^{\text{ori}}$). This divergence promotes a substantial shift in learned features by disrupting the alignment between the model's feature representation and any backdoor features, particularly in neurons identified as suspicious.

\begin{figure*}[t]
  \centering
  \includegraphics[width=0.88\linewidth]{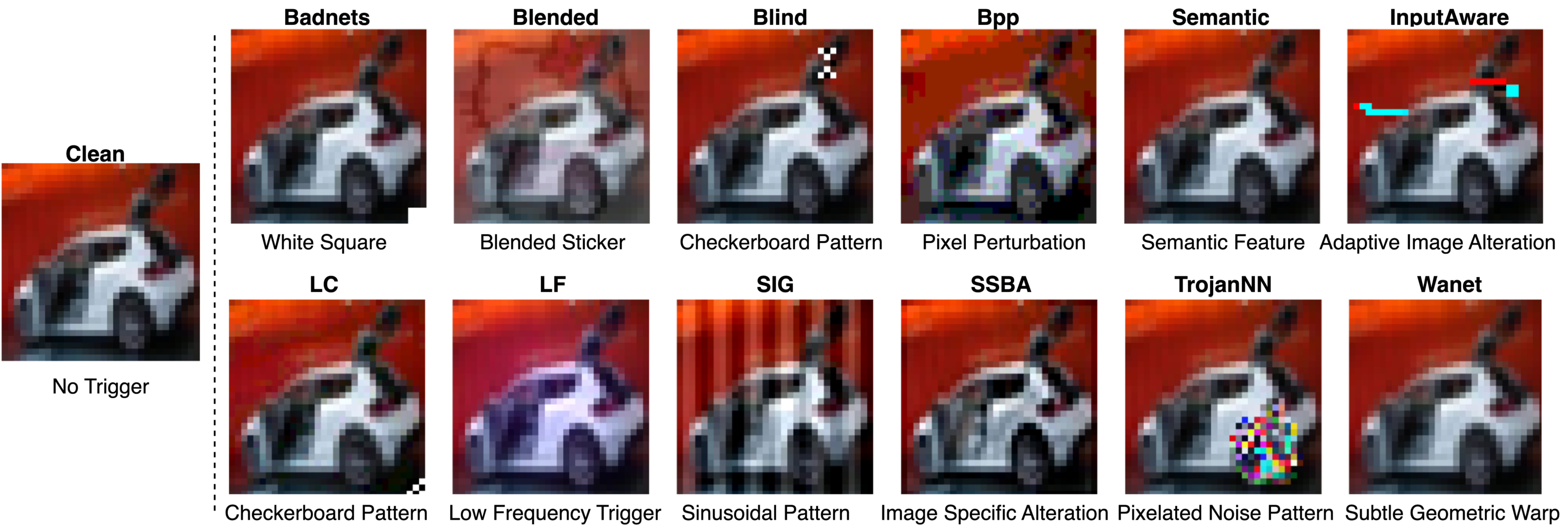}
  \caption{Example images of 12 backdoor samples from CIFAR-10. The caption below each figure is the backdoor trigger.}
  \label{fig: compiled_backdoor}
  
\end{figure*}

Cosine similarity, chosen for its ability to measure vector orientation independently of magnitude, supports this by offering a refined measure of divergence. This metric, focusing on the cosine of the angle between vectors, is crucial for recalibrating neuron weights where backdoor influences typically manifest in orientation rather than magnitude. It effectively addresses the challenges posed by high-dimensional spaces in neural networks, enabling the disentanglement and correct orientation of neuron weight vectors that may have been manipulated to encode backdoor triggers. The inclusion of cosine similarity regularization presents several key advantages:

\begin{itemize}
    \item \textbf{Targeted Adjustment:} Cosine similarity facilitates precise neuron weight adjustments by modifying orientation while preserving magnitude to ensure that defense efforts are focused on effectively disrupting the backdoor.
    
    \item \textbf{Preservation of Model Utility:} By adjusting orientation, it maintains the model’s clean performance without affecting its overall parameter distribution.
    
    \item \textbf{Robustness to Varied Attack Strategies:} Focusing on orientation allows the model to counter diverse backdoor strategies, even those that subtly manipulate weights.
\end{itemize}

\section{Experiments}
\label{Sec:Experiment}
In this section, we thoroughly evaluate the effectiveness of ULRL by conducting a series of comprehensive experiments. 

\subsection{Experimental Setup}
We evaluated our approach on four datasets: 1) CIFAR-10~\cite{krizhevsky2009learning} with 50,000 training and 10,000 test 32$\times$32 color images across 10 classes; 2) GTSRB~\cite{6033395} with 39,200 training and 12,600 test variable-resolution traffic sign images across 43 classes; 3) CIFAR-100~\cite{krizhevsky2009learning} with 50,000 training and 10,000 test images across 100 categories; and 4) Tiny-ImageNet~\cite{chrabaszcz2017downsampled} with 100,000 training and 10,000 validation 64$\times$64 images across 200 classes. We used three architectures: PreAct-ResNet18~\cite{he2016identity}, VGG19-BN~\cite{simonyan2015deep}, and ViT-B-16~\cite{dosovitskiy2021imageworth16x16words}.

\noindent \textbf{Attack Setup.} We consider a comprehensive set of 12 kinds of backdoor attacks: BadNets~\cite{tian2019badnet}, Blended~\cite{xin2017targeted}, Blind~\cite{Bagdasaryan2021Blind}, BPPAttack (BPP)~\cite{wang2022bppattack}, Label Consistent (LC)~\cite{https://doi.org/10.48550/arxiv.1912.02771}, Sinusoidal Signal (SIG)~\cite{barni2019new}, Low Frequency (LF)~\cite{Zeng_2021_ICCV}, Sample-Specific Backdoor Attack (SSBA)~\cite{li2021invisible}, Input-Aware Dynamic Backdoor (InputAware)~\cite{NEURIPS2020_234e6913}, Semantic~\cite{pmlr-v108-bagdasaryan20a}, Trojanning attack (TrojanNN)~\cite{yin2018trojaning}, and Warping-Based Poisoned Networks (WaNet)~\cite{nguyen2021wanet}. Figure~\ref{fig: compiled_backdoor} shows one sample of each attack. For consistency and fairness, we use the default settings from BackdoorBench~\cite{backdoorbench}, including trigger patterns, hyperparameters, and a universal 10\% poisoning ratio. The attack target is label 0, except for the semantic backdoor, which misclassifies images based on natural semantic features (e.g., green cars as frogs) without using artificial triggers. Specifically, semantic backdoor misclassifies images of green cars (class 2) as frogs (class 6). It should be noted that some attacks were excluded from GTSRB, CIFAR-100, and Tiny-ImageNet either because they were not designed for these datasets or had an ASR below 10\%. Additional details are in Appendix~\ref{Appendix: Attack Details}.

\noindent \textbf{Defense Setup.} We evaluate five state-of-the-art backdoor defense methods: Fine-pruning (FP)~\cite{DBLP:journals/corr/abs-1805-12185}, Adversarial Unlearning of Backdoors via Implicit Hypergradient (I-BAU)~\cite{DBLP:journals/corr/abs-2110-03735}, Adversarial Neuron Perturbation (ANP)~\cite{NEURIPS2021_8cbe9ce2}, Reconstructive Neuron Pruning (RNP)~\cite{li2023reconstructive}, and Feature Shift Tuning (FST)~\cite{min2023towards}. Each defense is limited to 1\% of clean samples per benchmark dataset. We also systematically evaluated the impact of the size of the defense data in one of the experiments. All methods follow BackdoorBench~\cite{backdoorbench} implementations, except FST and RNP, where we use their original open-source versions for fair comparison. For ULRL, we used SGD for 20 epochs with a momentum of 0.9, a uniform unlearning rate $\kappa =$ 0.01, and a clean threshold of 0.2 across all datasets. Relearning involved dataset-specific adjustments: for PreAct-ResNet18, the learning rate $\eta$ was 0.005 for CIFAR-10 and 0.001 for GTSRB, CIFAR-100, and Tiny-ImageNet, with $\alpha$ set to 0.7 for CIFAR-10 and 0.001 for the others. For VGG19-BN, the $\eta$ was 0.1 and $\alpha$ was 1 for CIFAR-10, while for Tiny-ImageNet, the values were 0.005 and 0.5, respectively. MAD threshold (\(\tau\)) and hard threshold (HT) are set to 3.5 and 2 for all cases. Baseline configurations are in Appendix~\ref{Appendix: Defense Details}.

\noindent \textbf{Evaluation Metrics.} We evaluate ULRL using two metrics: C-ACC (model accuracy on clean samples) and ASR (the proportion of poisoned samples misclassified as the attacker's target label). An effective defense reduces ASR while maintaining C-ACC. C-ACC is measured on clean test samples, and ASR is assessed by applying the backdoor trigger to all test samples except those in the target class to check for misclassification.

\begin{table*}[htbp]
\caption{Performances of 6 backdoor defense methods against 12 backdoor attacks. The experiments were performed on CIFAR-10, GTSRB, CIFAR-100, and Tiny-ImageNet with PreAct-ResNet18 using only 1\% clean data. The best results are \textbf{boldfaced}.}
\label{table: Preact_Performance}
\renewcommand{\arraystretch}{1.05} 
\centering
\begin{adjustbox}{width=1.00\linewidth} 
\begin{tabular}{c|c|cc|cc|cc|cc|cc|cc|cc}
\toprule
\multirow{2}{*}{Dataset} & \multirow{2}{*}{\begin{tabular}[c]{@{}c@{}}Backdoor \\ Attacks\end{tabular}} & \multicolumn{2}{c|}{No Defense} & \multicolumn{2}{c|}{FP} & \multicolumn{2}{c|}{I-BAU} & \multicolumn{2}{c|}{ANP} & \multicolumn{2}{c|}{RNP} & \multicolumn{2}{c|}{FST} & \multicolumn{2}{c}{\textbf{ULRL (ours)}} \\ \cline{3-16} 
 &  & ACC & ASR & ACC & ASR & ACC & ASR & ACC & ASR & ACC & ASR & ACC & ASR & ACC & ASR \\ \midrule
\multirow{11}{*}{\shortstack{CIFAR\\-10}} 
& BadNets    & 91.32 & 95.03  & \textbf{-0.87}  & 1.14   & -12.76 & 1.87  & -8.41  & 0.02   & -11.15 & \textbf{0.00}   & -2.11  & 0.16  & -2.36  & 2.70  \\ 
& Blended    & 93.47 & 99.92  & -1.75  & 9.95   & -19.10 & 10.06 & -6.50  & 5.28   & -23.06 & \textbf{0.00}   & -1.72  & 11.50 & \textbf{-1.64}  & 0.54  \\
& Blind      & 83.32 & 99.94  & \textbf{+9.79}  & \textbf{0.00}   & +1.30  & 83.22 & -0.40  & 99.92  & 0.00   & 73.77  & +8.65  & \textbf{0.00}  & +8.93  & \textbf{0.00}  \\
& BPP        & 91.00 & 100.00 & +0.67  & 38.67  & -1.24  & 98.12 & -6.59  & 0.68   & -4.75  & 1.80   & \textbf{+1.04}  & 0.72  & +0.86  & \textbf{0.09}  \\
& LC         & 84.59 & 99.92  & +3.05  & 31.37  & \textbf{+5.39}  & 75.98 & -1.24  & 0.01   & -8.29  & \textbf{0.00}   & +3.97  & 15.07 & +4.23  & 11.56 \\
& SIG        & 84.48 & 98.27  & +3.26  & 30.49  & -13.69 & \textbf{0.00}  & -6.20  & 0.04   & -7.84  & 0.01   & +2.94  & \textbf{0.00}  & \textbf{+3.60}  & \textbf{0.00}  \\
& LF         & 93.19 & 99.28  & -2.07  & 29.61  & -32.59 & 87.66 & -3.99  & \textbf{0.06}   & -7.80  & 0.30   & -1.73  & 32.49 & \textbf{-1.64}  & 0.24  \\
& SSBA       & 92.88 & 97.86  & -2.79  & 16.67  & -13.38 & 4.24  & -7.36  & \textbf{0.48}   & -15.10 & 1.72   & \textbf{-2.09}  & 4.66  & -2.19  & 0.66  \\
& Semantic   & 91.06 & 100.00 & -1.25  & 60.00  & -10.74 & 20.00 & -0.20  & 60.00  & -4.98  & \textbf{0.00}   & \textbf{0.00}   & 20.00 & -0.23  & \textbf{0.00}  \\
& InputAwa & 90.67 & 98.26  & +1.29  & 14.47  & -0.69  & 70.41 & +0.39  & 0.65   & +0.06  & 0.70   & \textbf{+1.40}  & 0.08  & +0.79  & \textbf{0.00}  \\
& TrojanNN   & 93.42 & 100.00 & -1.89  & 41.99  & -25.43 & 98.22 & -8.98  & 0.11   & -29.47 & \textbf{0.00}   & \textbf{-1.75}  & 51.70 & -2.72  & \textbf{0.00}  \\
& WaNet      & 91.25 & 89.73  & \textbf{+1.23}  & 30.52  & -2.40  & 54.14 & -9.00  & 0.24   & -2.95  & 0.93   & +0.52  & 0.86  & +0.29  & \textbf{0.06}  \\ \midrule
& Average    & 90.05 & 98.18  & +0.72  & 25.41  & -10.44 & 50.33 & -4.87  & 13.96  & -9.61  & 6.60   & \textbf{+0.76}  & 11.44 & +0.66  & \textbf{1.23}  \\  \hline
 \multirow{9}{*}{GTSRB}
& BadNets    & 97.62 & 95.48  & \textbf{+0.76}  & \textbf{0.00}   & -27.55 & 26.72 & -1.25  & \textbf{0.00}   & -3.69  & 1.61   & -5.82  & 0.08  & +0.14  & \textbf{0.00}  \\
& Blended    & 98.62 & 100.00 & \textbf{-1.60}  & 89.73  & -90.60 & 97.69 & -5.74  & 5.86   & -8.64  & 14.00  & -13.19 & 59.90 & -1.66  & \textbf{0.00}  \\
& Blind      & 42.56 & 69.29  & \textbf{+55.98} & 25.94  & +14.33 & \textbf{0.00}  & +2.52  & 38.15  & -11.06 & 53.75  & +33.39 & \textbf{0.00}  & +54.45 & 22.83 \\
& BPP        & 97.43 & 99.90  & \textbf{+1.37}  & 0.04   & -4.09  & 44.78 & +0.97  & \textbf{0.00}   & -11.67 & \textbf{0.00}   & -5.57  & \textbf{0.00}  & +0.40  & \textbf{0.00}  \\
& LF         & 97.89 & 99.36  & -1.32  & 93.16  & -67.46 & 92.76 & -6.96  & \textbf{0.00}   & \textbf{-0.11}  & 0.78   & -14.45 & 0.02  & -1.71  & \textbf{0.00}  \\
& SSBA       & 97.90 & 99.47  & -1.35  & 73.06  & -68.79 & \textbf{0.00}  & -7.69  & \textbf{0.00}   & -7.16  & 37.51  & -11.99 & 5.32  & \textbf{-1.11}  & \textbf{0.00}  \\
& InputAwa & 98.76 & 95.92  & \textbf{+0.42}  & 0.03   & -4.94  & 8.61  & -0.05  & 0.65   & -3.32  & \textbf{0.00}   & -4.33  & \textbf{0.00}  & -1.09  & \textbf{0.00}  \\
& TrojanNN   & 98.57 & 100.00 & -2.13  & 70.89  & -51.33 & 60.25 & -1.51  & \textbf{0.00}   & \textbf{-1.41}  & 13.35  & -13.24 & 1.32  & -2.09  & \textbf{0.00}  \\
& WaNet      & 97.74 & 94.25  & \textbf{+0.47}  & 6.88   & -11.33 & 33.54 & +0.01  & \textbf{0.00}   & -1.64  & \textbf{0.00}   & -4.39  & 0.34  & -0.71  & \textbf{0.00}  \\ \midrule
& Average    & 91.90 & 94.85  & \textbf{+5.84}  & 39.97  & -34.64 & 40.48 & -2.19  & 4.96   & -5.41  & 13.44  & -4.40  & 7.44  & +5.18  & \textbf{2.54}  \\  \hline
  \multirow{8}{*}{\shortstack{CIFAR\\-100}} 
& BadNets    & 67.23 & 87.43  & -6.16  & \textbf{0.00}   & -8.63  & \textbf{0.00}  & -5.52  & \textbf{0.00}   & -5.73  & \textbf{0.00}   & -22.62 & 0.01  & \textbf{-1.35}  & \textbf{0.00}  \\
& Blended    & 69.28 & 95.95  & -13.50 & 40.03  & -11.60 & 89.67 & -4.71  & 83.34  & -25.36 & 3.19   & -24.67 & \textbf{0.00}  & \textbf{-2.18}  & \textbf{0.00}  \\
& Blind      & 43.92 & 100.00 & \textbf{+26.19} & 100.00 & +13.39 & 76.44 & -0.01  & 100.00 & \textbf{0.00}   & 100.00 & +2.58  & \textbf{0.00}  & +24.60 & \textbf{0.00}  \\
& BPP        & 64.01 & 98.88  & +0.11  & 0.12   & -0.23  & 99.46 & -5.85  & \textbf{0.00}   & -20.70 & \textbf{0.00}   & -18.53 & \textbf{0.00}  & \textbf{+0.96}  & \textbf{0.00}  \\
& SSBA       & 69.07 & 97.22  & -14.31 & 11.48  & -15.59 & 78.09 & \textbf{+21.14} & \textbf{0.00}   & -2.01  & 0.02   & -23.73 & \textbf{0.00}  & -2.10  & \textbf{0.00}  \\
& InputAwa & 65.24 & 98.63  & -1.31  & 8.67   & -3.65  & 94.89 & -5.57  & 0.01   & -2.24  & 0.03   & -19.32 & 0.02  & \textbf{+0.67}  & \textbf{0.00}  \\
& TrojanNN   & 69.82 & 100.00 & -15.00 & 70.10  & -18.57 & 98.22 & -6.77  & 67.75  & -12.93 & \textbf{0.00}   & -24.71 & 3.86  & \textbf{-2.68}  & \textbf{0.00}  \\
& WaNet      & 64.05 & 97.73  & +0.11  & 45.46  & -2.13  & 38.89 & -5.76  & \textbf{0.00}   & -1.54  & 0.03   & -18.59 & \textbf{0.00}  & \textbf{+0.19}  & \textbf{0.00}  \\ \midrule
& Average    & 64.08 & 96.98  & -2.98  & 34.48  & -5.88  & 71.96 & -1.63  & 31.39  & -8.81  & 12.91  & -18.70 & 0.49  & \textbf{+2.26}  & \textbf{0.00}  \\  \hline
  \multirow{9}{*}{\shortstack{Tiny-\\Image\\Net}} 
& BadNets    & 55.94 & 100.00 & -7.57  & 0.32   & -6.59  & 99.50 & -4.73  & 2.41   & -1.58  & 0.01   & -30.77 & 1.81  & \textbf{-0.89}  & \textbf{0.00}  \\
& Blended    & 55.81 & 99.71  & -9.20  & 96.72  & -6.76  & 92.56 & \textbf{-0.59}  & 75.41  & -20.20 & \textbf{0.00}   & -31.66 & 10.24 & -0.97  & \textbf{0.00}  \\
& BPP        & 58.15 & 100.00 & -11.70 & 1.95   & -7.84  & 89.75 & -3.75  & 99.08  & -32.10 & \textbf{0.00}   & -34.39 & 1.10  & \textbf{-1.80}  & \textbf{0.00}  \\
& LF         & 55.48 & 98.61  & -9.51  & 89.13  & -5.66  & 97.97 & -4.29  & 93.57  & -10.52 & \textbf{0.00}   & -31.88 & 7.65  & \textbf{-0.77}  & \textbf{0.00}  \\
& SSBA       & 55.15 & 97.71  & -9.18  & 84.26  & -6.86  & 92.15 & -2.60  & 83.66  & -8.08  & 0.21   & -32.12 & 2.85  & \textbf{-1.33}  & \textbf{0.00}  \\
& InputAwa & 57.78 & 98.04  & -6.43  & 0.14   & -5.93  & 39.09 & -3.82  & 0.90   & -29.29 & \textbf{0.00}   & -37.11 & 3.88  & \textbf{-4.92}  & \textbf{0.00}  \\
& TrojanNN   & 55.75 & 99.98  & -8.90  & 84.59  & -6.58  & 99.43 & -0.14  & 2.23   & -16.42 & \textbf{0.00}   & -30.39 & 0.46  & \textbf{-0.85}  & \textbf{0.00}  \\
& WaNet      & 56.59 & 99.49  & -7.84  & 7.46   & -5.41  & 96.84 & -3.69  & 6.16   & -3.74  & \textbf{0.00}   & -33.03 & 0.24  & \textbf{-1.91}  & \textbf{0.00}  \\ \midrule
& Average    & 56.33 & 99.19  & -8.79  & 45.57  & -6.45  & 88.41 & -2.95  & 45.43  & -15.24 & 0.03   & -32.67 & 3.53  & \textbf{-1.68}  & \textbf{0.00}  \\
 \bottomrule 
\end{tabular}
\end{adjustbox}
\end{table*}

\begin{table*}[htbp]
\caption{Performances of 6 backdoor defenses against 9 backdoor attacks. The experiments were performed on CIFAR-10, GTSRB, CIFAR-100, and Tiny-ImageNet with VGG19-BN using only 1\% clean defense data. The best results are \textbf{boldfaced}.}
\label{table: VGG19_Performance} 
\renewcommand{\arraystretch}{1.05} 
\centering
\begin{adjustbox}{width=1.00\linewidth} 
\begin{tabular}{c|c|cc|cc|cc|cc|cc|cc|cc}
\toprule
\multirow{2}{*}{Dataset} & \multirow{2}{*}{\begin{tabular}[c]{@{}c@{}}Backdoor \\ Attacks\end{tabular}} & \multicolumn{2}{c|}{No Defense} & \multicolumn{2}{c|}{FP} & \multicolumn{2}{c|}{I-BAU} & \multicolumn{2}{c|}{ANP} & \multicolumn{2}{c|}{RNP} & \multicolumn{2}{c|}{FST} & \multicolumn{2}{c}{\textbf{ULRL (ours)}} \\ \cline{3-16} 
 &  & ACC & ASR & ACC & ASR & ACC & ASR & ACC & ASR & ACC & ASR & ACC & ASR & ACC & ASR \\ \midrule
\multirow{9}{*}{\shortstack{CIFAR\\-10}} 
& BadNets    & 90.42 & 94.43  & \textbf{-1.09} & 33.86         & -5.04  & 8.46          & -8.13          & 2.69          & -8.39          & \textbf{0.00} & -5.51          & 43.30         & -7.11          & 7.79          \\
& Blended    & 91.91 & 99.50  & \textbf{-1.27} & 94.45         & -11.67 & 32.30         & -4.46          & 8.63          & -19.52         & \textbf{0.00} & -1.31          & 96.12         & -5.43          & 9.51          \\
& Blind      & 88.67 & 99.29  & \textbf{+2.91} & 5.54          & -4.61  & 42.86         & -7.20          & 99.01         & 0.00           & 99.29         & +2.91          & 46.08         & +1.36          & \textbf{2.00} \\
& BPP        & 89.31 & 99.79  & \textbf{+1.23} & 4.19          & -1.14  & 89.68         & -8.04          & \textbf{0.01} & -1.71          & 0.53          & +0.11          & 3.34          & -0.08          & 2.53          \\
& LC         & 83.24 & 74.58  & \textbf{+4.73} & 48.70         & +1.19  & 40.50         & -4.26          & \textbf{0.00} & -2.26          & 0.04          & +2.40          & 37.93         & +1.66          & 12.82         \\
& SIG        & 83.48 & 98.87  & \textbf{+3.75} & 7.10          & -7.41  & 0.28          & -0.98          & \textbf{0.00} & +2.57          & 0.02          & +1.64          & 2.54          & +0.31          & 6.18          \\
& InputAwa & 89.70 & 97.02  & \textbf{+2.26} & 14.47         & -1.99  & 52.14         & -0.83          & 0.68          & -1.99          & \textbf{0.50} & -0.45          & 82.74         & -0.79          & 12.47         \\
& TrojanNN   & 91.56 & 100.00 & \textbf{-1.04} & 7.72          & -13.16 & 57.74         & -1.91          & \textbf{0.51} & -23.68         & 0.97          & -1.73          & 12.20         & -6.59          & 10.60         \\
& WaNet      & 84.58 & 96.49  & +5.69          & 1.17          & +3.89  & 2.31          & +0.71          & \textbf{0.15} & +4.48          & 0.63          & \textbf{+5.82} & 53.27         & +2.13          & 25.72         \\ \midrule
& Average & 88.37 & 95.51 & \textbf{+1.91} & 24.13 & -4.44  & 36.25 & -3.90 & 11.36 & -5.61 & 11.33 & +0.43  & 41.95         & -1.62 & \textbf{9.96}         \\ \hline 			                                 					
 \multirow{7}{*}{GTSRB}
& BadNets    & 97.28 & 93.44  & +0.36          & \textbf{0.00} & -43.02 & \textbf{0.00} & -0.24          & \textbf{0.00} & -1.56          & \textbf{0.00} & -10.91         & \textbf{0.00} & \textbf{+0.48} & \textbf{0.00} \\
& Blended    & 97.74 & 100.00 & +0.05          & 100.00        & -11.43 & 99.60         & -2.81          & 38.42         & \textbf{+0.08} & 99.76         & -9.84          & \textbf{0.00} & -1.53          & 34.21         \\
& Blind      & 89.84 & 8.70   & \textbf{+7.98} & \textbf{0.00} & -8.89  & \textbf{0.00} & -0.53          & 7.25          & -24.85         & 17.48         & -4.16          & \textbf{0.00} & +7.52          & \textbf{0.00} \\
& BPP        & 97.82 & 63.61  & \textbf{+0.71} & \textbf{0.00} & -7.61  & 1.92          & +0.08          & \textbf{0.00} & -11.57         & \textbf{0.00} & -20.92         & \textbf{0.00} & -1.09          & \textbf{0.00} \\
& InputAwa & 96.86 & 97.46  & \textbf{+0.68} & 0.14          & -4.32  & 25.69         & -0.02          & \textbf{0.00} & -13.04         & \textbf{0.00} & -15.66         & \textbf{0.00} & -0.24          & \textbf{0.00} \\
& TrojanNN   & 97.77 & 100.00 & -0.10          & 100.00        & -13.76 & 0.57          & -0.13          & \textbf{0.00} & -6.31          & \textbf{0.00} & -12.98         & \textbf{0.00} & \textbf{+0.08} & \textbf{0.00} \\
& WaNet      & 94.76 & 98.32  & \textbf{+4.01} & 0.23          & -0.79  & \textbf{0.00} & +3.38          & \textbf{0.00} & +2.72          & \textbf{0.00} & -15.05         & \textbf{0.00} & +2.43          & \textbf{0.00} \\ \midrule
& Average & 97.15 & 93.18 & \textbf{+0.95} & 33.40 & -13.49 & 21.30 & +0.04 & 6.40  & -4.95 & 16.63          & -14.23 & \textbf{0.00} & +0.02 & 5.70          \\ \hline
 \multirow{6}{*}{\shortstack{CIFAR\\-100}} 
& BadNets    & 61.27 & 88.32  & \textbf{-0.36} & 19.96         & -8.13  & \textbf{0.00} & -5.65          & \textbf{0.00} & -7.97          & \textbf{0.00} & -17.20         & \textbf{0.00} & -1.02          & \textbf{0.00} \\
& Blended    & 64.25 & 98.98  & \textbf{-1.90} & 89.08         & -8.11  & 75.39         & -4.85          & 36.93         & -13.88         & \textbf{0.00} & -15.32         & \textbf{0.00} & -2.45          & \textbf{0.00} \\
& BPP        & 60.13 & 98.28  & \textbf{+2.74} & 0.13          & -1.10  & 98.90         & -0.39          & \textbf{0.00} & -0.03          & 0.01          & -33.76         & \textbf{0.00} & +1.07          & \textbf{0.00} \\
& InputAwa & 59.41 & 97.55  & \textbf{+3.33} & 0.33          & -4.29  & 75.48         & -28.75         & 49.19         & -0.14          & 18.86         & -32.97         & \textbf{0.00} & +2.60          & \textbf{0.00} \\
& TrojanNN   & 64.89 & 99.98  & -2.50          & 99.59         & -11.35 & 24.00         & -6.29          & 1.37          & -21.36         & \textbf{0.00} & -16.10         & \textbf{0.00} & \textbf{-1.89} & \textbf{0.00} \\
& WaNet      & 57.91 & 96.22  & \textbf{+4.18} & 0.04          & -0.87  & 23.08         & +1.15          & 0.04          & +1.92          & \textbf{0.00} & -36.88         & \textbf{0.00} & +2.59          & \textbf{0.00} \\ \midrule
& Average & 61.49 & 96.43 & \textbf{+0.92} & 34.86 & -5.64  & 49.48 & -7.46 & 14.59 & -6.91 & 3.15           & -25.37 & \textbf{0.00}          & +0.15 & \textbf{0.00} \\  \hline    		
  \multirow{4}{*}{\shortstack{Tiny-\\Image\\Net}} 
& BadNets    & 43.56 & 99.96  & \textbf{+3.48} & 99.86         & -2.09  & 99.83         & -2.74          & \textbf{0.00} & +0.38          & \textbf{0.00} & -13.08         & \textbf{0.00} & +2.95          & \textbf{0.00} \\
& Blended    & 51.21 & 99.33  & \textbf{-3.01} & 96.14         & -7.53  & 92.87         & -4.56          & 98.73         & -7.08          & 97.63         & -12.37         & 84.75         & -3.48          & \textbf{0.00} \\
& BPP        & 55.34 & 99.96  & -2.08          & 0.49          & -8.38  & 97.28         & \textbf{-0.17} & \textbf{0.00} & -3.09          & \textbf{0.00} & -12.34         & 0.38          & -3.13          & 5.87          \\
& InputAware & 53.55 & 99.87  & -1.17          & 1.47          & -7.64  & 55.10         & -0.20          & 0.80          & \textbf{-0.06} & \textbf{0.00} & -12.96         & 0.11          & -1.18          & \textbf{0.00} \\\midrule
& Average    & 50.92 & 99.78  & \textbf{-0.70} & 49.49         & -6.41  & 86.27         & -1.92          & 24.88         & -2.46          & 24.41         & -12.69         & 21.31         & -1.21          & \textbf{1.47}  \\ 
 \bottomrule
\end{tabular}
\end{adjustbox}
\end{table*}

\begin{figure*}[t]
  \centering
  \includegraphics[width=0.76\linewidth]{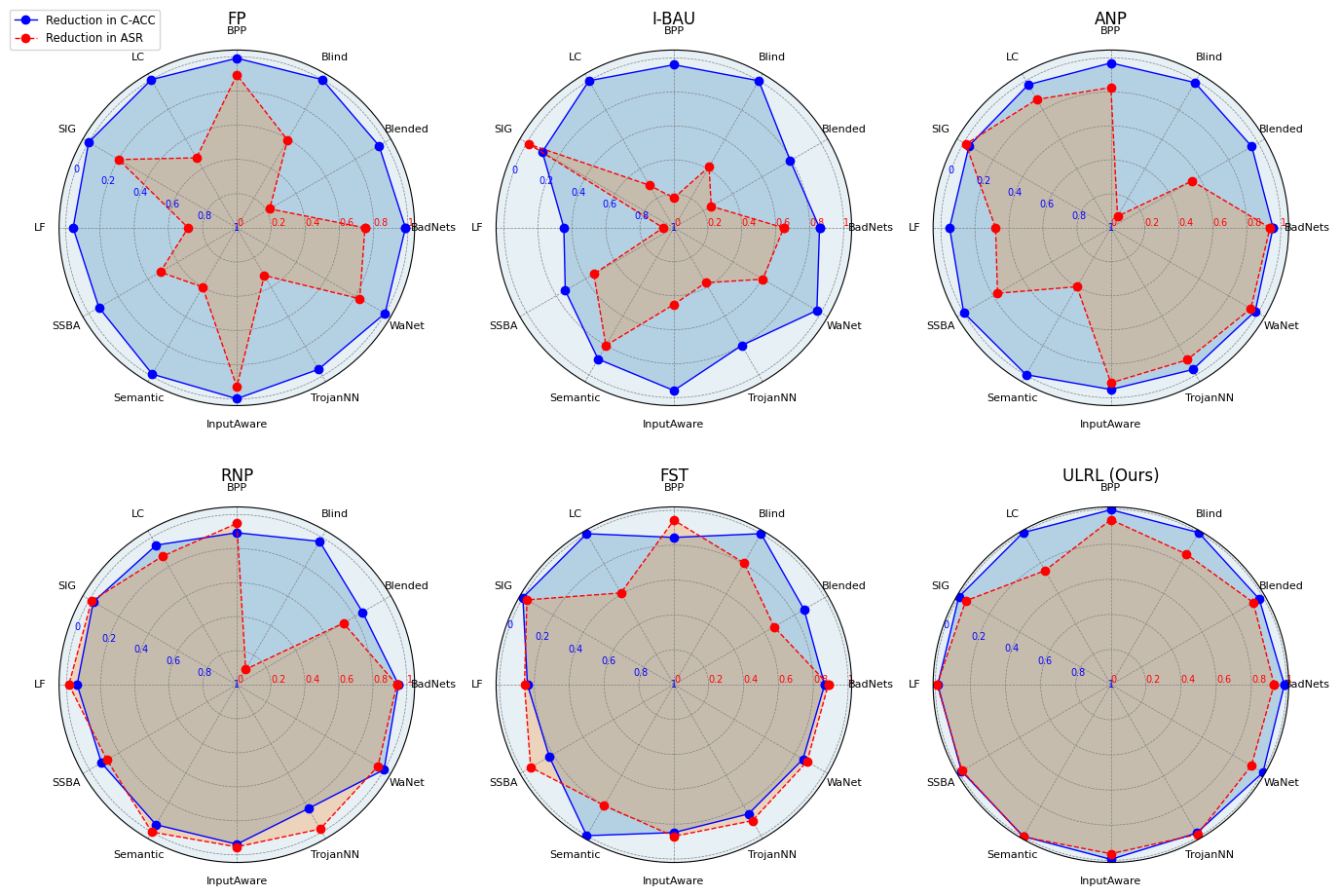}
  \caption{Impact of 6 Defenses on C-ACC and ASR Across 12 Attack Types Averaged Over 4 Datasets with 2 Architectures.}
  \label{fig: defense_comparison}
\end{figure*}

\subsection{Experimental Results}
\label{subsection: main_defense_result}
In the following, we answer five research questions.

\noindent \textbf{\textit{RQ1 (Backdoor Mitigation and Performance Maintenance): Can ULRL mitigate different backdoor attacks?}} To answer this, we evaluated ULRL against various backdoors. Tables~\ref{table: Preact_Performance} and~\ref{table: VGG19_Performance} show ULRL’s C-ACC and ASR results for PreAct-ResNet18 and VGG19-BN architectures. Our findings demonstrate that ULRL significantly reduces ASR across all backdoor attacks, effectively neutralizing them.

The average ASR reduction is 94.94\% for PreAct-ResNet18 and 90.46\% for VGG19-BN. In terms of C-ACC, PreAct-ResNet18 sees an average increase of 1.61\%, while VGG19-BN experiences a slight decrease of 0.73\%, with individual results varying. Notably, C-ACC against the Blind attack significantly improves across both architectures, demonstrating ULRL's effectiveness. In contrast, attacks such as Blended and BadNets show a slight decrease in C-ACC.

Furthermore, LC and Blind on PreAct-ResNet18 show higher residual ASRs, indicating greater resistance to ULRL. However, results from RQ3 show that increasing the defense data to 2\% neutralizes the ASR of LC while maintaining C-ACC. For the Blind attack on GTSRB, a higher learning rate $\eta =$ 0.005 fully mitigates the attack. In VGG19-BN, increasing the $\eta =$ 0.2 for CIFAR-10 and $\eta =$ 0.005 for GTSRB eliminates residual ASRs from LC, Blended, and InputAware attacks with minimal accuracy loss.

ULRL also accurately identifies suspicious neurons (e.g., $6^{th}$ for Semantic and $0^{th}$ for others) in most cases, except for the Blind. Blind leverages advanced optimization and evasion strategies without direct access to training data, making it harder to detect. However, ULRL still effectively mitigates it by neutralizing backdoor-influenced neurons, demonstrating its precision in identifying compromised neural targets.

\noindent \textbf{\textit{RQ2: How does ULRL perform compared to baseline defenses?}} Our comparative analysis, summarized in Tables~\ref{table: Preact_Performance} and ~\ref{table: VGG19_Performance}, demonstrates that ULRL outperforms five baseline defenses by effectively maintaining a high C-ACC and significantly reducing ASR in complex attack scenarios. On CIFAR-10, ULRL not only eliminates ASR but also improves C-ACC in the case of Blind. The results for GTSRB are similarly impressive, with high C-ACC preservation and ASR reduced to almost zero, even under complex attacks with very few exceptions. On CIFAR-100 and Tiny-ImageNet, despite the complexities, ULRL's efficacy persists, reducing ASR to near-zero across all attacks, demonstrating remarkable versatility.

Each defense exhibits unique strengths and weaknesses, often dependent on the attack's complexity. FP is typically effective against straightforward attacks but struggles against more sophisticated triggers, where it often sacrifices C-ACC to lower ASR. I-BAU adapts well to clearly defined patterns and dynamically adjusts its parameters, but it suffers from consistency issues and can overly compromise generalization. ANP is beneficial in situations where malicious neurons are identifiable, though it also faces trade-offs between C-ACC and ASR, especially in complex scenarios. RNP attempts a balanced approach, moderately impacting C-ACC while trying to reduce ASR, yet it does not always effectively mitigate sophisticated attacks. FST excels in scenarios involving conspicuous feature anomalies but encounters challenges with more nuanced attacks, where backdoor triggers are subtle and well-integrated into benign features. While these methods have applicable scenarios and benefits, ULRL demonstrates superior adaptability and robustness, maintaining high C-ACC and drastically reducing ASR across various datasets and attacks. 

\begin{figure*}[t]
   \label{fig:defense_size}
  \centering
  \includegraphics[width=1\linewidth]{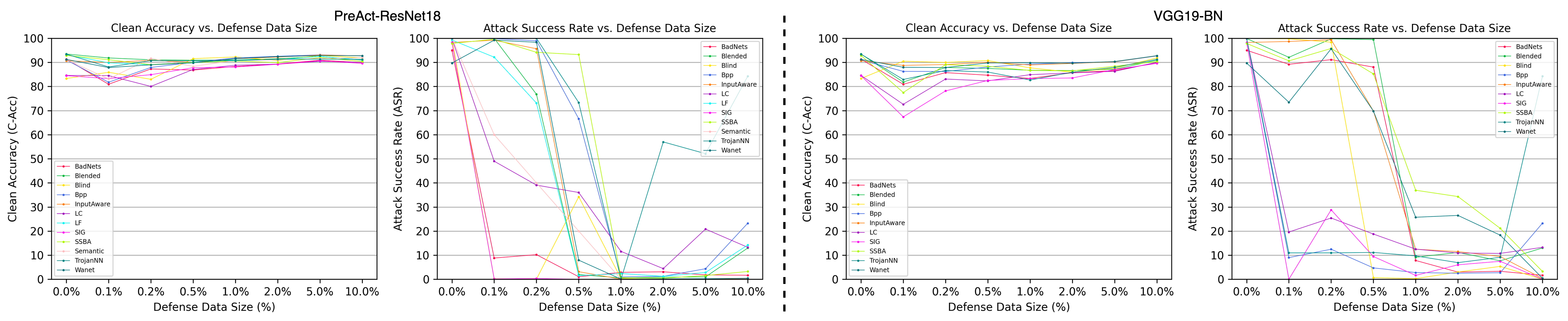}
  \caption{Defense performance against various backdoor attacks by data size on CIFAR-10 for PreAct-ResNet18 and VGG19-BN.}

\end{figure*}

Figure~\ref{fig: defense_comparison} illustrates the overall effectiveness of 6 defenses on two architectures. The blue and red lines quantify the reductions in C-ACC and ASR. Optimal defenses are indicated by blue lines near the edge, suggesting a negligible impact on C-ACC, and red lines close to the edge, reflecting a high reduction in ASR. FP offers a defense with moderate ASR reduction and minimal effect on C-ACC. I-BAU effectively reduces ASR in certain attacks, such as SIG and Semantic, but significantly compromises C-ACC. ANP and RNP exhibit a trade-off between mitigating ASR and preserving C-ACC, with ANP providing a moderate compromise on both fronts and RNP achieving considerable ASR reduction at a notable cost to C-ACC, particularly for BPP, Blended, and TrojanNN. FST stands out for its substantial ASR reduction but comes at the expense of a higher impact on C-ACC, especially for InputAware, TrojanNN, and WaNet. In contrast, ULRL is the most effective, achieving the highest reduction in ASR with the negligible impact on C-ACC. ULRL's superior performance in various attack contexts confirms its effectiveness. 

\begin{table*}[t]
\caption{Average Running Time for Defenses on PreAct-ResNet18 and VGG19-BN (in seconds).}
\label{tab:runtime}
\centering
\begin{adjustbox}{width=1\linewidth} 

\begin{tabular}{l|c|c|c|c|c|c||c|c|c|c|c|c}
\toprule
& \multicolumn{6}{c||}{PreAct-ResNet18} & \multicolumn{6}{c}{VGG19-BN} \\
\cmidrule(lr){2-7} \cmidrule(lr){8-13}
Dataset & FP & IBAU & ANP & RNP & FST & \textbf{ULRL (Ours)} & FP & IBAU & ANP & RNP & FST & \textbf{ULRL (Ours)} \\
\midrule
CIFAR-10 & 797.87 & 38.55 & 613.90 & 124.10 & 24.21 & 80.38 & 912.23 & 47.94 & 416.31 & 200.10 & 26.94 & 77.51 \\
GTSRB & 851.43 & 52.04 & 657.92 & 224.44 & 24.27 & 110.76 & 924.00 & 48.94 & 461.65 &  231.07 & 27.24 & 92.59 \\
CIFAR-100 & 686.13 & 51.23 & 563.00 & 158.92 & 24.18 & 97.61 & 804.56 & 48.42 & 412.66 &  202.93 & 26.87 & 78.86 \\
Tiny-ImageNet & 1184.41 & 219.50 & 1682.41 & 277.46 & 41.57 & 165.25 & 1186.10 & 127.57 & 1086.12 & 303.15 & 35.68 & 135.43 \\
\bottomrule
\end{tabular}
\end{adjustbox}
\end{table*}

\noindent \textbf{\textit{RQ3: How does the size of clean defense data affect the results?}} \label{RQ3: Clean Size} In practice, users often rely on a limited set of clean data to conduct an in-house evaluation of third-party trained models due to the high costs associated with building larger datasets. Therefore, the mitigation methods must remain effective even with minimal data. To this end, we analyzed the performance of ULRL with varying proportions of the CIFAR-10 defense dataset, precisely 0.1\% (50), 0.2\% (100), 0.5\% (250), 1\% (500), 2\% (1000), 5\% (2500) and 10\% (5000) of the clean training set. The results are presented in Figure~\ref{fig:defense_size}.

ULRL initially displayed variability in defense efficacy with minimal clean data (0.1\% and 0.2\%) on both architectures. A moderate decline in average C-ACC signified an adaptation phase, where ULRL began addressing backdoor influences. A marked decrease in ASR illustrated ULRL's aptitude for leveraging even scarce defense data. However, with extremely limited datasets, ASR reduction was inadequate for more sophisticated attacks such as LF, WaNet, SSBA, and InputAware. We conjecture that effective unlearning requires a critical amount of clean data; with insufficient data, pinpointing the backdoor neuron, which is crucial for successful relearning, becomes less effective, thus only marginally reducing ASR. Enhancing the defense data to 0.5\% saw a discernible improvement in ULRL's defensive capabilities, characterized by a substantial decrease in ASR for both architectures, confirming the method's effectiveness against various backdoor attacks with minimal data. As the defense data size increased to the optimal range of 1\% to 2\%, we observed the most striking balance: a significant diminishment in ASR, minimized variability in defense efficacy, and, in certain cases, enhancements in C-ACC. For instance, within this range, ULRL improved the PreAct-ResNet18 model's C-ACC by 8.93\% for Blind and 3.60\% for SIG attacks, with ASR effectively neutralized. Concurrently, the VGG19-BN model's C-ACC increased by 1.36\% for Blind and 2.13\% for WaNet attacks.

The nuanced nature of this balance became more evident when expanding the defense data size to 5\% and beyond. Here, both architectures began to show signs of fluctuations in effectiveness, suggesting a suboptimal utilization of clean data, and at 10\%, specific attacks like TrojanNN and BPP demonstrated a plateau in defense gains, indicating a limit to the benefits of additional clean data.

Through comparative analysis, we identified 1\% to 2\% clean data as consistently effective for ULRL across both architectures. This range significantly reduces ASR while preserving accuracy for various backdoor attacks. However, complex attacks like TrojanNN and LC reveal challenges in creating a universal defense. Overall, ULRL proves to be an effective defense mechanism, emphasizing the importance of optimizing clean defense data.
\begin{table}[t]
    \caption{ULRL Performance on ViT-B-16 Architecture.}
    \label{tab:vit-b-16-results}
    \centering
    \begin{adjustbox}{width=0.89\linewidth}
    \begin{tabular}{l|l|cc|cc}
        \hline
        \multirow{2}{*}{Dataset} & \multirow{2}{*}{\begin{tabular}[c]{@{}c@{}}Backdoor \\ Attacks\end{tabular}} & \multicolumn{2}{c|}{No Defense} & \multicolumn{2}{c}{\textbf{ULRL (ours)}} \\
        \cline{3-6}
        & & C-ACC & ASR & C-ACC & ASR \\
        \midrule
        \multirow{6}{*}{CIFAR-10} 
        & BadNets & 94.78 & 93.77 & -5.37 & 1.09 \\
        & Blended & 96.54 & 99.68 & -3.06 & 0.08 \\
        & BPP & 91.38 & 99.37 & -1.77 & 0.91 \\
        & SIG & 86.84 & 92.77 & -1.69 & 0.04 \\
        & SSBA & 96.17 & 97.81 & -2.33 & 0.16 \\
        & WaNet & 89.12 & 80.95 & +1.40 & 0.99 \\\midrule
        & Average & 92.47 & 94.06 & -2.14 & 0.55 \\ 
        \bottomrule
    \end{tabular}
    \end{adjustbox}
\end{table}

\noindent \textbf{\textit{RQ4: How efficient is our approach?}} To answer this, we analyzed the computational efficiency of ULRL on PreAct-ResNet18 and VGG19-BN using an NVIDIA RTX 3090 GPU, an Intel Xeon Platinum 8350C CPU, and 42GB of RAM with PyTorch. Table~\ref{tab:runtime} shows the average running times of ULRL, which are consistent across both architectures. For PreAct-ResNet18, ULRL took 80.38, 110.76, 97.61, and 165.25 seconds on CIFAR-10, GTSRB, CIFAR-100, and Tiny-ImageNet, respectively. For VGG19-BN, the corresponding times were 77.51, 92.59, 78.86, and 135.43 seconds. 

In terms of computational complexity, the per-iteration cost of ULRL is comparable to that of standard gradient-based optimization methods. In the unlearning phase, ULRL performs gradient ascent on the clean defense dataset $\mathcal{D}_d$, and convergence is ensured by a stopping criterion based on $CA_{\min}$. Although this phase introduces additional iterations, our experiments indicate that it converges within a modest number of epochs. In the relearning phase, ULRL employs SGD with momentum to minimize a composite loss that includes a cosine similarity regularization term; this term adds only negligible overhead relative to the overall gradient computation. Overall, ULRL's combined computational complexity remains competitive with methods such as FP and IBAU, while being significantly more efficient than approaches like ANP that require extensive per-class optimization. These results highlight that ULRL not only achieves strong mitigation effectiveness and accuracy but also offers practical computational efficiency for resource-constrained environments.

\noindent \textbf{\textit{{RQ5:} Can ULRL effectively mitigate backdoor attacks against the Vision Transformer?}} We evaluated ULRL on the ViT-B-16 architecture using CIFAR-10, testing a range of representative backdoor attacks, including BadNets, Blended, BPP, SIG, SSBA, and WaNet. These attacks were chosen for their diversity, providing a thorough evaluation. As shown in Table~\ref{tab:vit-b-16-results}, ULRL significantly reduces ASR across all attacks, lowering ASR to below 1\% in most cases while maintaining C-ACC between 92.47\% and 90.33\%. Notably, for WaNet, ULRL not only reduces ASR but also improves C-ACC by 1.40\%. On average, ULRL reduces ASR from 94.06\% to 0.55\%, with only a 2.14\% decrease in C-ACC. The average running time is 780 seconds. These results demonstrate ULRL’s effectiveness in mitigating backdoors on ViT-B-16 while preserving accuracy.

\begin{table}[t]
    \caption{Ablation study results for ULRL method across various backdoor attacks on CIFAR-10 with PreActResNet18.}
    \label{tab:ablation-study}
    \centering
    \begin{adjustbox}{width=0.92\linewidth} 
    \begin{tabular}{l|cc|cc|cc}
        \toprule
        \multirow{2}{*}{\begin{tabular}[c]{@{}c@{}}Backdoor \\ Attacks\end{tabular}}  & \multicolumn{2}{c|}{No Defense} & \multicolumn{2}{c|}{Full} & \multicolumn{2}{c}{Without Reg} \\
        \cline{2-7}
         & C-ACC & ASR & C-ACC & ASR & C-ACC & ASR \\
        \midrule 	 				
        BadNets      & 91.32 & 95.03 & -2.36 & 2.70 & -3.22 & 22.26   \\
        Blended     & 93.47 & 89.13 & -1.64 & 0.54 & -3.15 & 96.54    \\
        SIG         & 84.48 & 01.40 & +3.60 & 0.00 & +3.49 & 28.06    \\
        WaNet       & 91.25 & 89.73  & +0.29 & 0.06 & +0.70 & 86.09   \\
        \bottomrule
    \end{tabular}
    \end{adjustbox}
\end{table}

\section{Ablation Study}
\label{Sec:Ablation}
In this section, we conduct multiple ablation studies to understand the role played by different components in ULRL. 

\subsection{Regularization Term}
We highlight the importance of the regularization term during retraining by comparing ULRL with and without the cosine similarity regularization. Table~\ref{tab:ablation-study} shows that removing the regularization significantly increases ASR across all attacks, underscoring its critical role in ULRL’s effectiveness (see Appendix~\ref{sec:alpha_sensitivity} for a sweep of the regularization weight~$\alpha$).

\subsection{Reinitializing Random Neurons}
ULRL reinitializes up to 2 neurons, achieving lower residual ASR and higher C-ACC than random selection. To test if random selection yields similar results, Table~\ref{tab:random-results} shows the impact of reinitializing 1, 2, 5, and 10 randomly selected neurons. Reinitializing 1 neuron results in a high residual ASR and reduced C-ACC, while increasing the number lowers ASR but causes a greater accuracy drop. This demonstrates that ULRL effectively balances neuron reinitialization to minimize ASR while maintaining performance.

\subsection{MAD Threshold for Identifying Suspicious Neurons}
The MAD threshold \(\tau\) is key to identifying suspicious neurons during unlearning. A low \(\tau\) flags too many neurons, reducing performance, while a high \(\tau\) may miss some, allowing backdoor effects to persist. We tested \(\tau\) values from 2 to 4 in 0.5 increments and analyzed their effects on ASR and C-ACC. Table~\ref{tab:mad-results} shows that ULRL effectively reduces ASR across all values. Lower \(\tau\) over-penalizes the model, reducing C-ACC, while higher values (\(\tau = 3.5\) and 4) maintain low ASR and better preserve C-ACC. \(\tau = 3.5\) consistently achieves the best balance. These results confirm that ULRL’s performance remains robust for $\tau$ values between 2 and 4.

\begin{table*}[t]
    \caption{Impact of Reinitializing Random Neurons Against BadNets on CIFAR-10 with PreAct-ResNet18.}
    \label{tab:random-results}
    \centering
    \begin{adjustbox}{width=0.92\linewidth}
    \setlength{\tabcolsep}{6pt}
    \begin{tabular}{l|cc|cc|cc|cc|cc|cc}
        \hline
        \multirow{2}{*}{Attack} &
        \multicolumn{2}{c|}{No Defense} & \multicolumn{2}{c|}{ULRL} & \multicolumn{2}{c|}{1 Neuron} & \multicolumn{2}{c|}{2 Neurons} & \multicolumn{2}{c|}{5 Neurons} & \multicolumn{2}{c}{10 Neurons} \\
        \cline{2-13}
        & C-ACC & ASR & C-ACC & ASR & C-ACC & ASR & C-ACC & ASR & C-ACC & ASR & C-ACC & ASR \\
        \midrule
        BadNets & 91.32 & 95.03 & -2.36 & 2.70 & -3.4 & 14.28 & -4.72 & 2.67 & -12.2 & 2.10 & -12.2 & 0.53 \\ 
        Blended & 93.47 & 89.13 & -1.64 & 0.54 & -1.94 & 66.26 & -1.93 & 57.78 & -3.43 & 2.77 & -3.46 & 0.04 \\ 
        WaNet & 91.25 & 89.73 & +0.29 & 0.06 & +0.18 & 62.91 & -7.92 & 45.57 & -19.35 & 28.24 & -32.36 & 0.05 \\  
        \bottomrule
    \end{tabular}    
    \end{adjustbox} 
\end{table*}

\begin{table*}[t]
    \caption{MAD Threshold \(\tau\) Effect Against BadNets Attack on PreAct-ResNet18 Architecture.}
    \label{tab:mad-results}
    \centering
    \begin{adjustbox}{width=0.92\linewidth}
    \setlength{\tabcolsep}{6pt}
    \begin{tabular}{l|cc|cc|cc|cc|cc|cc}
        \hline
        \multirow{2}{*}{Dataset} &
        \multicolumn{2}{c|}{No Defense} & \multicolumn{2}{c|}{\(\tau\)=2} & \multicolumn{2}{c|}{\(\tau\)=2.5} & \multicolumn{2}{c|}{\(\tau\)=3} & \multicolumn{2}{c|}{\(\tau\)=3.5} & \multicolumn{2}{c}{\(\tau\)=4} \\
        \cline{2-13}
        & C-ACC & ASR & C-ACC & ASR & C-ACC & ASR & C-ACC & ASR & C-ACC & ASR & C-ACC & ASR \\
        \midrule
        CIFAR-10 & 91.32 & 95.03 & -2.36 & 2.70 & -2.36 & 2.70 & -2.36 & 2.70  & -2.36 & 2.70 & -2.36 & 2.70 \\
        GTSRB & 97.62 & 95.48 & -0.05 & 0.00 & -0.05 & 0.00 & +0.13 & 0.00 & +0.13 & 0.00 & +0.13 & 0.00 \\
        CIFAR-100 & 67.23 & 87.43 & -10.72 & 0.00 & -5.98 & 0.00 & -4.19 & 0.00 & -3.41 & 0.00 & -1.39 & 0.05 \\
        Tiny-ImageNet & 55.94 & 100.00 & -7.00 & 0.00 & -4.75 & 0.00 & -3.43 & 0.00 & -2.88 & 0.00 & -2.54 & 0.07 \\
        \bottomrule
    \end{tabular}
    \end{adjustbox}
\end{table*}

\begin{table*}[t]
    \caption{Hard Threshold (HT) Effect Against BadNets Attack on PreAct-ResNet18 Architecture.}
    \label{tab:ht-results}
    \centering
    \begin{adjustbox}{width=0.92\linewidth}
    \setlength{\tabcolsep}{6pt}
    \begin{tabular}{l|cc|cc|cc|cc|cc|cc}
        \hline
        \multirow{2}{*}{Dataset} &
        \multicolumn{2}{c|}{No Defense} & \multicolumn{2}{c|}{HT=1} & \multicolumn{2}{c|}{HT=2 (default)} & \multicolumn{2}{c|}{HT=3} & \multicolumn{2}{c|}{HT=5} & \multicolumn{2}{c}{HT=10} \\
        \cline{2-13}
        & C-ACC & ASR & C-ACC & ASR & C-ACC & ASR & C-ACC & ASR & C-ACC & ASR & C-ACC & ASR \\
        \midrule
        CIFAR-10 & 91.32 & 95.03 & -2.36 & 2.70 & -2.36 & 2.70 & N/A & N/A & N/A & N/A & N/A & N/A \\ 
        GTSRB & 97.62 & 95.48 & +0.14 & 0.00 & +0.14 & 0.00 & -0.36 & 0.00 & N/A & N/A & N/A & N/A \\
        CIFAR-100 & 67.23 & 87.43 & -0.59 & 1.09 & -1.35 & 0.00 & -2.21 & 0.00 & -4.02 & 0.00 &  -7.21 & 0.00 \\
        Tiny-ImageNet & 55.94 & 100.00 & -0.46 & 0.00 & -0.89 & 0.00 & -1.62 & 0.00 & -4.53 & 0.00 & -4.53 & 0.00 \\
        \bottomrule
    \end{tabular}
    \end{adjustbox}
\end{table*}

\subsection{Hard Thresholds for Suspicious Neurons} In addition to MAD, ULRL uses a hard threshold (HT) to select neurons with strong outliers. We tested HT values on the BadNets attack across CIFAR-10, GTSRB, CIFAR-100, and Tiny-ImageNet. The thresholds were 1 and 2 neurons for all datasets, 3 neurons for GTSRB, CIFAR-100, and Tiny-ImageNet, and 5 and 10 neurons specifically for CIFAR-100 and Tiny-ImageNet. Table~\ref{tab:ht-results} shows that higher HT values reduce ASR but significantly decrease C-ACC for complex datasets. Lower HT values (1 or 2) maintain low ASR with minimal C-ACC impact across all datasets. HT = 2 provides the best balance between C-ACC and backdoor mitigation.

\noindent \textbf{Note:} ``N/A'' entries indicate that the tested hard threshold (HT) values are not applicable for specific datasets. ULRL identifies up to 20\% neurons as suspicious depending on the data set and the type of attack. For CIFAR-10, 20\% corresponds to at most 2 neurons, making HT values greater than 2 inapplicable. Similarly, for GTSRB, up to 3 neurons can be flagged, restricting the tested HT values to 1, 2, and 3, with higher thresholds being irrelevant.

\section{Related Work}
\label{Sec:Related}

\noindent \textbf{Backdoor Attacks.} Backdoor attacks mislead models into abnormal behavior on trigger-stamped samples while behaving normally on benign ones. They generally fall into two categories: data poisoning and model poisoning. Data poisoning manipulates training data using various triggers based on visibility~\cite{li2020invisible, liu2020reflection}, locality~\cite{shafahi2018poison, barni2019new}, additivity~\cite{xin2017targeted, Zeng_2021_ICCV} and specificity~\cite{tian2019badnet,li2021invisible}. Model poisoning allows attackers to jointly optimize trigger and model weights~\cite{Doan2021lira, xin2017targeted, nguyen2021wanet}. 

Traditional methods like BadNets~\cite{tian2019badnet} and Blended~\cite{xin2017targeted} use explicit triggers, while LF~\cite{Zeng_2021_ICCV} and SIG~\cite{barni2019new} employ subtler strategies, making triggers less noticeable. SIG uses a sinusoidal signal as its trigger, blending with image properties to create label-consistent backdoors. Blind attacks~\cite{Bagdasaryan2021Blind} dynamically poison inputs during training by manipulating the loss function without needing access to training data. Advanced methods like BPP~\cite{wang2022bppattack} use image quantization and dithering to imperceptibly embed triggers, while TrojanNN~\cite{yin2018trojaning} inverts neural networks to embed a universal trojan trigger. LC~\cite{https://doi.org/10.48550/arxiv.1912.02771} and SSBA~\cite{li2021invisible} deeply embed backdoors in models, making detection difficult. InputAware~\cite{NEURIPS2020_234e6913} and WaNet~\cite{nguyen2021wanet} employ adaptive learning and warping functions to create customized poisoned inputs, enhancing attack effectiveness and evasion. These methods show how attackers can improve their techniques to avoid detection and increase effectiveness. 

\noindent \textbf{Backdoor Defenses.} Backdoor defenses protect DNNs by either detecting or removing backdoors. Detection strategies analyze prediction biases or statistical deviations in feature space~\cite{li2021rethinking, DBLP:journals/corr/abs-1811-00636, DBLP:journals/corr/abs-1811-03728, Liu_2022_CVPR}, while advanced methods use reverse engineering for trigger detection~\cite{wang2019neural, DBLP:journals/corr/abs-2006-05646, DBLP:journals/corr/abs-1902-06531, hu2022trigger}. Removal strategies eliminate backdoors through robust training or post-processing purification. Robust training~\cite{huang2022backdoor, li2021antibackdoor, DBLP:journals/corr/abs-2110-03735, min2023towards} aims to prevent backdoor learning but may reduce accuracy and increase costs. Post-processing methods, like pruning and unlearning~\cite{DBLP:journals/corr/abs-1805-12185, wang2019neural, NEURIPS2021_8cbe9ce2, zheng2022datafree, foret2021sharpnessaware, li2023reconstructive}, mitigate compromised neurons but can affect accuracy and struggle to generalize across architectures. Innovative defenses, such as attention, channel analysis, and value estimation, offer more effective backdoor removal~\cite{DBLP:journals/corr/abs-2101-05930, zheng2022datafree, guan2022fewshot}. Techniques like Adversarial Unlearning (I-BAU)\cite{DBLP:journals/corr/abs-2110-03735} and Adversarial Neuron Pruning (ANP)\cite{NEURIPS2021_8cbe9ce2} target adversarially sensitive components, with varying success depending on attack complexity.

Backdoored DNNs contain a mix of clean and backdoor neurons, with the latter activated by trigger patterns. Identifying and pruning these neurons can significantly purify the model~\cite{NEURIPS2021_8cbe9ce2}. Reconstructive Neuron Pruning (RNP)~\cite{li2023reconstructive} uses asymmetric neuron-level unlearning, but balancing pruning effectiveness with clean accuracy remains challenging, often leading to trade-offs between defense and performance. Feature Shift Tuning (FST)~\cite{min2023towards} mitigates data poisoning backdoor attacks by incorporating a classification loss penalty based on the divergence between the classifier's adjusted and original weights. However, FST requires full reinitialization of classifier neuron parameters, risking the loss of valuable feature representations. Its focus on inner product regularization emphasizes weight magnitude but neglects directional properties, potentially reducing effectiveness. Additionally, the projection constraint to prevent parameter explosion limits the model’s exploration and adds complexity to fine-tuning.

Our ULRL mitigates vulnerable neurons in the classifier layer through precise interventions, overcoming the drawbacks of current defenses. By focusing on the classifier, ULRL simplifies the defense process, avoiding the complexities of extensive pruning or stringent thresholds. Unlike FST, which relies on full reinitialization and feature shift regularization, ULRL introduces relearning to preserve learned features. It prioritizes minimizing cosine similarity between weights, focusing on directional alignment rather than magnitude, eliminating the need for norm constraints and streamlining optimization. This approach enhances model generalization and provides robust defense against complex backdoor strategies, avoiding the limitations of projection constraints in existing methods.

\section{Limitation}
\label{Sec:Limitation}
ULRL's effectiveness decreases with very limited clean data, requiring at least 1\% for optimal performance, which is still modest compared to other methods. ULRL mitigates backdoors by reinitializing suspicious neurons but may miss some if many neurons are affected, potentially degrading performance. Attackers could exploit this by targeting more neurons, though targeting the top two suspicious neurons remains effective. If all last-layer neurons are compromised, MAD may fail, but reinitializing all last-layer neurons can mitigate backdoors, though it may slow optimization.

\section{Conclusion}
\label{Sec:Conclusion}
We introduce ULRL, a novel method to protect DNNs from backdoor attacks by adjusting vulnerable neurons. ULRL maintains high clean accuracy and significantly reduces ASR across various datasets, architectures, and attacks. Future work will extend ULRL to other network types and backdoor threats, including data-free settings. Our open-source code is available at \url{https://github.com/NayMyatMin/ULRL}. 

\section*{Acknowledgments}
This research is supported by the Ministry of Education, Singapore under its Academic Research Fund Tier 3 (Award ID: MOET32020-0004).

{\appendices
\section{Implementation Details}
\subsection{Attack Details}
\label{Appendix: Attack Details}

This section outlines the configuration for 12 attacks on CIFAR-10, CIFAR-100, GTSRB, and Tiny-ImageNet datasets with consistent hyperparameters. All-to-one attacks mislabel poisoned samples as class 0. Models were optimized using SGD (momentum 0.9, weight decay $5e^{-4}$, batch size 128). CIFAR-10 and CIFAR-100 models were trained for 100 epochs, GTSRB for 50, and Tiny-ImageNet for 200. Cosine Annealing Learning Rate was used for CIFAR and GTSRB, except for controllable attacks, while Tiny-ImageNet used ReduceLROnPlateau. Initial learning rates were 0.1 (CIFAR and GTSRB) and 0.01 (WaNet and Tiny-ImageNet).

All backdoor attacks followed standard configurations from BackdoorBench~\cite{backdoorbench}, except for the Semantic. \textbf{BadNets} used a 3x3 white checkerboard pattern on the bottom right. The \textbf{Blended} attack used a `Hello Kitty' pattern with a 0.2 alpha. \textbf{Label Consistent} applied PGD with a 1.5 step size over 100 steps, epsilon 8, and delta 40. \textbf{Blind} balanced normal and backdoor losses (1.0 and 0.9) over 1000 batches. \textbf{BPP} used a 0.1 negative ratio and squeeze number of 8. \textbf{SIG} employed a sinusoidal signal (amplitude 40, frequency 6). \textbf{LF} had a fooling rate of 0.2 with 50 termination iterations. \textbf{SSBA} encoded a single bit, with a training step at $14e^{4}$. \textbf{Input-Aware} applied a 0.01 learning rate with a MultiStepLR schedule. \textbf{TrojanNN} targeted two linear neurons, optimizing an `apple' logo over 1000 iterations. \textbf{WaNet} used a grid rescale and cross-ratio of 2. \textbf{Semantic} backdoors misclassified green cars as Frogs with a learning rate of 0.01 over 100 epochs~\cite{pmlr-v108-bagdasaryan20a}.

\subsection{Defense Details}
\label{Appendix: Defense Details}
We evaluated ULRL through conducted comparative analyses against five defense methods, following configurations from BackdoorBench and other benchmark sources.

\begin{itemize}
    \item \textbf{FP:} Implemented with an initial learning rate of 0.01 over 100 epochs, pruning targeted the last convolutional layer until C-ACC dropped below 80\%. 
    \item \textbf{ANP:} Perturbation budget was 0.4 and the trade-off coefficient 0.2. Thresholds from 0.4 to 0.9 were tested to balance low ASR and high C-ACC.
    \item \textbf{RNP:} Using the open-source implementation, the unlearning phase ran for 20 epochs (learning rate 0.01, weight decay $5e^{-2}$) until C-ACC reached 10\%, followed by a 20-epoch recovery with dynamic threshold pruning within [0.4, 0.7] interval and a 0.2 learning rate.
    \item \textbf{FST:} FST was implemented using its open-source code. SGD was set with a learning rate of 0.01 and momentum of 0.9 for CIFAR-10 and GTSRB, and $1e^{-3}$ for CIFAR-100 and Tiny-ImageNet. FST ran for 10 epochs on CIFAR-10 and 15 for the other datasets. The regularization parameter, $\alpha$, was set to 0.2 for CIFAR-10, 0.1 for GTSRB, and 0.001 for CIFAR-100 and Tiny-ImageNet, balancing model fidelity and robustness. For VGG19-BN, FST was applied to all final linear layers.
\end{itemize}

The batch size was standardized to 256 for all defenses, ensuring a fair and consistent evaluation of each method's effectiveness in mitigating backdoors.

\section{Effect of Clean Dataset Size}
We address the mitigation failures observed with very small clean datasets (0.1\%, 0.2\%, 0.5\%). Unlearning, essential for identifying and mitigating backdoor neurons, requires a sufficient amount of clean data for two primary reasons:

\begin{enumerate}
    \item \textbf{Identification of Suspicious Neurons}: ULRL detects neurons with abnormal activations or weight changes during unlearning, signaling backdoor influence. 
    \item \textbf{Statistical Robustness}: Larger datasets ensure reliable observations, minimizing false positives and negatives.
\end{enumerate}

\noindent \textbf{Impact of Extremely Small Dataset:} When the dataset size is extremely small (0.1\% to 0.2\%), the following issues arise:
\begin{itemize}
    \item \textbf{Limited Coverage}: A small dataset may not adequately represent the clean data's feature space.
    \item \textbf{Inadequate Statistical Power}: Fewer clean examples limit the ability to capture neuron activation variability, making it harder to reliably identify suspicious behavior.
\end{itemize}

Our results in RQ3 (Subsection \ref{RQ3: Clean Size}) show that Blended, LF, SSBA, InputAware, and TrojanNN attacks are difficult to mitigate with 0.1\% and 0.2\% clean data, but effectiveness improves significantly with 0.5\% and above. ASR decreases for most attacks, indicating better identification and mitigation of backdoors, while clean accuracy improves and stabilizes, demonstrating the model’s ability to maintain performance on legitimate tasks while effectively countering backdoor attacks.

\section{Empirical Comparison of Dispersion Measures for Suspicious Neuron Identification}
\label{Appendix: Measures}

In the unlearning phase of ULRL, we identify neurons exhibiting abnormal weight changes due to backdoor influences. We evaluated three statistical dispersion measures—Standard Deviation (SD), Interquartile Range (IQR), and Median Absolute Deviation (MAD). Our results demonstrate that MAD provides the most stable and accurate thresholding. Below, we present a comparative analysis, including a case study on the Blended attack with CIFAR-10 using PreAct-Resnet18.

\begin{enumerate}
\item{Standard Deviation (SD):}
SD is defined as: $\sigma = \sqrt{\frac{1}{N} \sum_{i=1}^{N} (x_i - \mu)^2},$ where \(x_i\) represents individual weight changes, and \(\mu\) is their mean. SD is highly sensitive to outliers, leading to inflated thresholds that fail to capture moderate yet significant deviations in neuron behavior.

\item{Interquartile Range (IQR):}
IQR is given by: IQR = $Q_3$ - $Q_1$, where $Q_1$ and $Q_3$ denote the 25th and 75th percentiles, respectively. While robust to extreme outliers, it often underestimates the spread of moderate deviations, resulting in overly conservative thresholds that miss key suspicious neurons.

\item{Median Absolute Deviation (MAD):}
MAD balances robustness and sensitivity, reliably distinguishing backdoor-affected neurons. Our experiments confirm that MAD effectively captures both extreme and moderate deviations, ensuring accurate detection of suspicious neurons.
\end{enumerate}

\textbf{Blended Attack on CIFAR-10 using PreAct-Resnet18.}
\begin{itemize}
    \item \textbf{SD:} Inflated by extreme weight updates, leading to an excessively high threshold that overlooks critical neurons.
    \item \textbf{IQR:} Too conservative, failing to detect moderate yet crucial deviations, missing key neurons. 
    \item \textbf{MAD:} Effectively flags suspicious deviations, precisely identifying backdoor-influenced neurons.
\end{itemize}

In summary, SD is overly sensitive to outliers, leading to unreliable thresholds. IQR is robust but too conservative, missing moderate deviations. However, MAD offers an optimal balance, providing stable, accurate thresholds. Given these results, we adopt MAD as the primary indicator in ULRL. 

\section{Comparison of Similarity Measures}
\label{Appendix:SimilarityComparison}

In our proposed ULRL method, the relearning objective is designed to achieve two goals: (i) preserve the model's clean task performance, and (ii) enforce a targeted divergence in the weight vectors of suspicious neurons relative to their original, potentially compromised states. As discussed in the main text, our approach employs cosine similarity as the regularization measure because it focuses solely on the orientation of the weight vectors, thereby decoupling directional changes from their magnitude. Cosine similarity is defined as:
${Cosine Similarity} = \frac{\boldsymbol{w}_i \cdot \boldsymbol{w}_i^{\text{ori}}}{\|\boldsymbol{w}_i\|_2 \|\boldsymbol{w}_i^{\text{ori}}\|_2},$
where \(\boldsymbol{w}_i\) is the current weight vector for neuron \(i\) and \(\boldsymbol{w}_i^{\text{ori}}\) is its original weight vector. This measure is inherently scale invariant, allowing it to capture changes in orientation independently of changes in magnitude. An alternative approach that has been used in prior works (e.g., in FST~\cite{min2023towards}) is to use inner product regularization, defined as: $\text{Inner Product} = \boldsymbol{w}_i \cdot \boldsymbol{w}_i^{\text{ori}}.$ While the inner product quantifies the overall alignment between the vectors, it conflates both magnitude and orientation. In our experiments, we found that directly using the inner product as a regularization measure did not preserve model utility as effectively as cosine similarity. Specifically, the inner product measure was more sensitive to changes in vector magnitude, which often led to overly aggressive adjustments that degraded clean accuracy. Moreover, it did not isolate the directional component of the weight changes; as a result, its ability to guide the relearning process to specifically disrupt backdoor influences was diminished.

\begin{table}[t]
\caption{Effect of the regularization weight $\alpha$ on PreAct‑ResNet18.}
\label{tab:alpha}
\centering
\begin{adjustbox}{width=1.0\linewidth}
\setlength{\tabcolsep}{2pt}
\begin{tabular}{l|cc|cc|cc|cc|cc}
\toprule
\multirow{2}{*}{Attack} &
\multicolumn{2}{c|}{NoDef.} &
\multicolumn{2}{c|}{$\alpha=0$} &
\multicolumn{2}{c|}{$\alpha=0.3$} &
\multicolumn{2}{c|}{$\alpha=0.7$} &
\multicolumn{2}{c}{$\alpha=1.0$} \\
\cline{2-11}
& ACC & ASR & ACC & ASR & ACC & ASR & ACC & ASR & ACC & ASR \\
\midrule
BadNets & 91.32 & 95.03 & $-3.22$ & 22.26 & $-2.39$ & 13.40 & $-2.36$ & 2.70 & $-5.39$ & 0.80 \\
WaNet     & 91.25 &  89.73 & $+0.70$ & 86.09 & $+0.51$ & 4.31 & $+0.29$ & 0.06 & $+0.11$ & 0.10 \\
\bottomrule
\end{tabular}
\end{adjustbox}
\vskip -0.1in
\end{table}

\noindent \textbf{Experimental Study.} To compare these two regularization measures, we conducted experiments on the CIFAR-10 on PreAct-Resnet18 using the Blended attack as a representative case. Two variants of relearning were implemented:
\begin{enumerate}
    \item \textbf{Cosine Similarity Regularization:} Our standard approach, where the regularization is based on cosine similarity between the updated and original weight vectors.
    \item \textbf{Inner Product Regularization:} An alternative variant where the cosine similarity term is replaced with the direct inner product \(\boldsymbol{w}_i \cdot \boldsymbol{w}_i^{\text{ori}}\) without normalization.
\end{enumerate}

We maintained all other hyperparameters constant between the two variants. The performance was evaluated in terms of the attack success rate (ASR) and clean accuracy (C-ACC).

\noindent \textbf{Results and Discussion.} Our experimental results indicate that the cosine similarity variant achieves a lower ASR and preserves clean accuracy more effectively compared to inner product variant. In the Blended attack case on CIFAR-10:
\begin{itemize}
    \item \textbf{Cosine Similarity:} The targeted regularization guided the suspicious neurons to diverge directionally from their compromised states without affecting the magnitude, thereby effectively neutralizing backdoor influences (0.54\%) and maintaining high C-ACC (91.83\%).  

    \item \textbf{Inner Product:} The inner product variant, by incorporating magnitude into its measure, led to less precise adjustments. This resulted in a higher residual ASR (14.51\%) and a modest degradation in C-ACC (84.23\%), indicating that the inner product measure did not preserve model utility as well.
\end{itemize}

The superior performance of cosine similarity can be attributed to its scale invariance and its exclusive focus on the directional changes in weight vectors. This property is critical in scenarios where backdoor triggers manipulate the orientation of neuron weights, as it ensures that the defense process concentrates on disentangling and rectifying these specific manipulations.

\section{Sensitivity to Regularization Weight \(\alpha\)}
\label{sec:alpha_sensitivity}

Table~\ref{tab:alpha} shows that with \(\alpha \ge 0.3\), ULRL significantly reduces the attack success rates (e.g., to 2.7\% for BadNets and 0.06\% for WaNet at \(\alpha=0.7\)) while maintaining clean accuracy with variations within 2 percentage points, demonstrating robust performance across a broad range of \(\alpha\) values.

\section{Impact of Lower Poisoning Rates}
\label{Appendix:LowPoisoning}

Attackers can typically use lower poisoning rates to remain stealthy, which presents a significant challenge for defenses. To evaluate the robustness of ULRL under these conditions, we conducted experiments on a PreAct-ResNet18 model under the Blended attack at various poisoning rates using CIFAR-10. Table~\ref{tab:low_poisoning} summarizes the performance results. As shown in Table~\ref{tab:low_poisoning}, the baseline model maintains high accuracy across poisoning rates, but ASR remains extremely high at 10\% and 5\% (above 99\%). At a 1\% poisoning rate, ASR decreases to 94.88\%, and further drops to 56.11\% at 0.1\% poisoning. This indicates that while the lower rates naturally weaken the backdoor signal, they still allow a substantial number of attacks to succeed if no defense is applied.

Our experiments with ULRL at a 10\% poisoning rate yield a purified model with 91.83\% clean accuracy and an ASR of only 0.54\%. Preliminary evaluations at lower poisoning rates (5\% and 1\%) indicate that ULRL continues to effectively mitigate backdoor effects, maintaining clean accuracy around 91.90–92.05\% while reducing the ASR to approximately 0.60\% and 2.10\%, respectively. Even at the extreme case of 0.1\% poisoning, ULRL significantly lowers the ASR relative to the baseline (14.50\% versus 56.11\%), though the defense challenge increases as the poisoning rate gets smaller. These results demonstrate that ULRL is robust across a range of poisoning rates. While our main experiments reported results at 10\% poisoning for consistency with existing benchmarks, the supplementary evaluation confirms that ULRL remains effective under more realistic, low-poisoning conditions. This underscores the practical applicability of our method in scenarios where attackers employ stealthy strategies.

\begin{table}[t]
\centering
\caption{Performance of ULRL at Different Poisoning Rates.}
\begin{adjustbox}{width=0.81\linewidth}
\setlength{\tabcolsep}{6pt}
\begin{tabular}{l|cc|cc}
\toprule
\multirow{2}{*}{Rate} & \multicolumn{2}{c|}{No Defense} & \multicolumn{2}{c}{ULRL} \\
\cmidrule(lr){2-5}
                                & ACC (\%) & ASR (\%)   & ACC (\%) & ASR (\%)       \\
\midrule
10\%                           & 93.47    & 99.92      & 91.83    & 0.54           \\
5\%                            & 93.67    & 99.61      & 91.90    & 0.60           \\
1\%                            & 93.76    & 94.88      & 92.05    & 2.10           \\
0.1\%                          & 93.80    & 56.11      & 92.10    & 14.50          \\
\bottomrule
\end{tabular}
\label{tab:low_poisoning}
\end{adjustbox}
\end{table}
}
\bibliography{IEEEabrv,main-cite}
\bibliographystyle{IEEEtran}

\end{document}